\documentclass{pasj01}

\usepackage{booktabs}
\usepackage{ifpdf}
\usepackage{pdfpages}
\usepackage{epstopdf}

\usepackage[]{natbib}

\Received{$\langle$2020.3.12$\rangle$}
\Accepted{$\langle$2020.5.13$\rangle$}

\begin{document}

\draft{}
\title{Cloud structures in M17 SWex : Possible cloud-cloud collision
}

\author{Shinichi W. \textsc{Kinoshita}\altaffilmark{1,2}, Fumitaka \textsc{Nakamura}\altaffilmark{1,2,3},
Quang \textsc{Nguyen-Luong}\altaffilmark{4,6},
Benjamin \textsc{Wu}\altaffilmark{2},
Tomomi \textsc{Shimoikura}\altaffilmark{5},
Koji \textsc{Sugitani}\altaffilmark{6},
Kazuhito \textsc{Dobashi}\altaffilmark{7},
Hideaki \textsc{Takemura}\altaffilmark{2,3},
Patricio \textsc{Sanhueza}\altaffilmark{2,3}, 
Kee-Tae \textsc{Kim}\altaffilmark{8,9}, 
Hyunwoo \textsc{Kang}\altaffilmark{8}, Neal J. \textsc{Evans}\altaffilmark{10} , Glenn J. \textsc{White}\altaffilmark{11,12}, Cassandra \textsc{Fallscheer}\altaffilmark{13}
}
\altaffiltext{1}{Department of Astronomy, the University of Tokyo, 7-3-1 Hongo Bunkyo, 113-0033 Tokyo, Japan}
\email{kinoshita.shinichi@nao.ac.jp}
\altaffiltext{2}{National Astronomical Observatory of Japan, NINS, 2-21-1 Osawa, Mitaka, Tokyo 181-8588, Japan}
\altaffiltext{3}{The Graduate University for Advanced Studies (SOKENDAI), 2-21-1 Osawa, Mitaka, Tokyo 181-0015, Japan}
\altaffiltext{4}{McMaster University, 1 James St N, Hamilton, ON, L8P 1A2, Canada}
\altaffiltext{5}{Faculty of Social Information Studies, Otsuma Womens University, Chiyoda-ku,Tokyo, 102-8357, Japan}
\altaffiltext{6}{Graduate School of Science, Nagoya City University, Mizuho-ku, Nagoya,467-8501, Japan}
\altaffiltext{7}{Department of Astronomy and Earth Sciences, Tokyo Gakugei University, 4-1-1, Nukuikitamachi, Koganei, Tokyo 184-8501, Japan}
\altaffiltext{8}{Korea Astronomy \& Space Science Institute, 776 Daedeokdae-ro, Yuseong-gu, Daejeon 34055, Republic of Korea}
\altaffiltext{9}{University of Science and Technology, Korea (UST), 217 Gajeong-ro, Yuseong-gu, Daejeon 34113, Republic of Korea}
\altaffiltext{10}{Department of Astronomy, The University of Texas at Austin, 2515 Speedway, Stop C1400, Austin, TX 78712-1205, USA}
\altaffiltext{11}{Department of Physics and Astronomy, The Open University, Walton Hall, Milton Keynes, MK7 6AA, UK }
\altaffiltext{12}{RAL Space, STFC Rutherford Appleton Laboratory, Chilton, Didcot, Oxfordshire, OX11 0QX, UK}
\altaffiltext{13}{Central Washington University, 400 E. University Way, Ellensburg, WA 98926 ,USA}

\KeyWords{ISM: kinematics and dynamics --- ISM: individual objects (the infrared dark cloud M17)
--- ISM: structure}

\maketitle

\begin{abstract}

Using wide-field $^{13}$CO ($J=1-0$) data taken with the Nobeyama 45-m telescope, 
we investigate cloud structures of the infrared dark cloud complex in M17 with SCIMES.
In total, we identified 118 clouds that contain 
11 large clouds with radii larger than 1 pc.
The clouds are mainly distributed in the two representative velocity ranges
of  10 $-$ 20 km s$^{-1}$ and 30 $-$ 40 km s$^{-1}$.
By comparing with the ATLASGAL catalog, we found that the majority of the $^{13}$CO clouds with 10 $-$ 20 km s$^{-1}$ and 30 $-$ 40 km s$^{-1}$ are likely located at distances of 2 kpc (Sagittarius arm) and 3 kpc (Scutum arm), respectively. 
Analyzing the spatial configuration of the identified clouds and their velocity structures, we attempt to reveal the origin of the cloud structure in this region. 
Here we discuss three possibilities: (1) overlapping with different velocities, (2) cloud oscillation, and (3) cloud-cloud collision.
From the position-velocity diagrams, we found spatially-extended faint emission between $\sim$ 20 km s$^{-1}$ and $\sim$ 35 km s$^{-1}$, which is mainly distributed in the spatially-overlapped areas of the clouds. Additionally, the cloud complex system is not likely to be gravitationally bound.
We also found that in some areas where clouds with different velocities overlapped, the magnetic field orientation changes abruptly.
The distribution of the diffuse emission in the position-position-velocity space and the bending magnetic fields appear to favor the cloud-cloud collision scenario compared to other scenarios.
In the cloud-cloud collision scenario,  
we propose that two $\sim$35 km s$^{-1}$ foreground clouds are colliding with clouds at $\sim$20 km s$^{-1}$ with a relative velocity of 15 km s$^{-1}$.  
These clouds may be substructures of two larger clouds having velocities of $\sim$ 35 km s$^{-1}$ ($\gtrsim 10^3 $ M$_{\odot}$) and $\sim$ 20 km s$^{-1}$
($\gtrsim 10^4 $ M$_{\odot}$), respectively.

\end{abstract}

\section{Introduction}
\label{sec:intro}

The high-mass stellar population dominates the energy budget of all forming stars, thus they significantly influence the evolution of the Galaxy \citep{Beuther_2007,Zinnecker_2007,Motte_2018}. Therefore, it is crucial to understand how high-mass stars form.
However, the formation process of high-mass stars is still under debate \citep{Bonnell_1998,Bonnell_2001,McKee_2003}. 
Recent observations have suggested that very massive and dense clumps in large clouds or infrared dark clouds (IRDCs) are promising sites of ongoing or future high-mass star formation \citep{Rathborne_2006,Sakai_2010,Shimoikura_2013,Sanhueza_2012,Sanhueza_2017,Contreras_2018,Sanhueza_2019}. 
Some theoretical and observational studies have proposed that collisions of clouds may trigger the formation of such massive and dense clumps and thus possibly trigger high-mass star formation
\citep{Scoville_1986,Tan_2000,Fukui_2014, Wu_2017,Dobashi_2019,Montillaud_2019,Kohno_2018}.  
In this paper, we 
present recent observations of the M17 region, focusing on the kinematics of the IRDC and potential scenarios that could explain them.
We note that \citet{Nishimura_2018} mentioned the possibility of a cloud-cloud collision around the M17 H{\scriptsize II} region, which is located next to the M17 IRDC region, including other possibilities.


\citet{Povich_2010} assumed that M17 SWex is located in the Sagittarius arm at a distance of $\sim$ 2.0 kpc which is  measured toward the M17 H{\scriptsize II} region
based on the parallax observations by \citet{Xu_2011},
because the LSR velocity of M17 SWex and the H{\scriptsize II}
region are almost the same (V$_{\rm LSR}$ $\simeq$ 20 km s$^{-1}$). On the other hand, on the basis of the VLBI observations toward a G14.33--0.64 core, located in the M17 SWex region, \citet{Sato_2010} derived a closer distance of 1.1 kpc $\pm$ 0.1 kpc.
They argued that the core belongs to the Sagittarius arm which is closer in the direction of M17 SWex. In contrast, \citet{Povich_2016} interpreted that this core might be a foreground object and the main structure of M17 SWex may be located to 2 kpc.  
Recently we analyzed the cloud structures in this area using the wide-field $^{12}$CO, $^{13}$CO, and C$^{18}$O observations \citep{Shimoikura_2019, Nguyen-Luong_2020}, and found that the molecular gas in the M17 and M17 SWex regions is continuous in the position-position velocity space, which is consistent with the argument of \citet{Povich_2016}. Thus, in this paper, we adopt a distance to M17 SWex of 2 kpc.

The M17 cluster has very high star density of $ > 10^3$ stars pc$^{-2}$, which is 10-100 times higher than 
that of the Orion Nebula Cluster, the nearest high-mass star forming region \citep{Lada_1991,Ando_2002}.
Previous molecular line observations have revealed that the H{\scriptsize II} region is surrounded by large molecular clouds
which are also sites of active ongoing high-mass star formation (hereafter the M17 H{\scriptsize II} region). 
\citet{Elmegreen_1979} discovered a significant amount of gas extending to 
the southwest direction from the M17 H{\scriptsize II} region, and Spitzer observations reveal that this cloud is dark in mid-infrared. The infrared dark feature suggests that high-mass star formation is at a very early stage of evolution without significantly disrupting the cloud (hereafter M17 SWex).
 
Recently, \citet{Povich_2010} and \citet{Povich_2016} found a very interesting characteristic of the protostellar populations 
in the M17 region.
From Spitzer observations, they discovered the mass function of protostars around the M17-H{\scriptsize II} region seems consistent with the Salpeter IMF, whereas that in M17 SWex is significantly steeper than the Salpeter IMF. 
In other words, the high-mass stellar population in M17 SWex is significantly deficient compared to what the Salpeter IMF expects.  It remains uncertain whether the future stellar mass function in the IRDC region will evolve into a Salpeter-like IMF, or if the high-mass stellar population stays deficient.
In either case, the M17 region is a feature-rich region that may shed light on what triggers or initiates high-mass star formation processes (see also \citet{Ohasi_2016} and \citet{Chen_2019} for recent ALMA observations in M17 SWex).

Recently, from wide field $^{12}$CO ($J=1-0$), $^{13}$CO ($J=1-0$), HCO$^+$ ($J=1-0$), and HCN$^+$ ($J=1-0$) observations, \citet{Nguyen-Luong_2020} investigated the cloud structure in the M17 region and found significant differences in dense gas distribution.  In the M17 H{\scriptsize II} region, about 27 \% of the molecular gas mass is concentrated in compact regions whose column densities are higher than 
$10^{23}$ cm$^{-2}$, or $\sim$ 1 g cm$^{-2}$, a theoretical threshold suggested for high-mass star formation proposed by \citet{Krumholz_2008}. In contrast,
in M17 SWex, the column densities are significantly lower than the threshold for high-mass star formation. Only 0.46\% of gas is denser than $10^{23}$ cm$^{-3}$.
In addition, the clumps in the M17 H{\scriptsize II} region are more massive than those in M17 SWex, but have similar sizes and therefore higher densities. Thus, \citet{Nguyen-Luong_2020} concluded that the deficiency in the high-mass stellar population in M17 SWex presumably comes from 
the extremely-low dense gas fraction. A similar trend is observed in other nearby star-forming regions such as Orion A and Aquila Rift (Nakamura et al. 2019).
The IRDC also contains prominent filamentary structures, some of which appear to be parallel to one another \citep{Busquet_2013,Busquet_2016}.

\citet{Sugitani_2019} performed wide-field near infrared polarization observations toward the M17 SWex region and found that 
the magnetic field is globally well ordered and perpendicular to the filamentary structures and the galactic plane.
From the $^{12}$CO ($J=3-2$) data, \citet{Sugitani_2019} also found two cloud components with different line-of-sight velocities: 
one at $\sim 20$ km s$^{-1}$, 
and the other at $\sim 35$ km s$^{-1}$. 
Regarding the cloud morphology, they suggested that a cloud-cloud collision might have happened along the direction of the ordered magnetic field.
The cloud-cloud collision is expected to promote the creation of high density clouds and clumps and 
thus possibly trigger high mass star formation.
One difficulty of this scenario is that the velocity of $\sim$ 35 km s$^{-1}$ is coincidentally 
close to the representative velocity of the adjacent Galactic spiral arm, the Scutum arm, at a distance of $\sim $ 3 kpc. The representative velocity of the Scutum arm is estimated to be 40 km s$^{-1}$ along the line of sight \citep{Nguyen-Luong_2020}.
Therefore, the $\sim$ 35 km s$^{-1}$ molecular gas component might simply overlap with the  
$\sim 20$ km s$^{-1}$ component along the line of sight.
In this paper, 
we attempt to constrain the cloud structure in a possible future high-mass star-forming region, M17 SWex, and analyze the possibility of the cloud-cloud collision scenario proposed by \citet{Sugitani_2019}.

The paper is organized as follows.
In section \ref{sec:obs}, we describe the details of the observations.  
We present the overall gas distribution in section \ref{sec:overall}.
Then, in section \ref{sec:result} we decompose clouds in the M17 SWex region with $^{13}$CO ($J=1-0$)
using SCIMES, which identifies continuous gas structures 
in the position-position-velocity space within dendrograms of emission
using the spectral clustering paradigm \citep{Colombo_2015}.
In section \ref{sec:ccc}, we discuss some possible scenarios which can explain observational  features in M17 SWex.
Finally, we summarize the main results in section \ref{sec:summary}.

\section{Observations}
\label{sec:obs}

\subsection{$^{12}$CO $(J=1-0)$ and $^{13}$CO $(J=1-0)$ observations}
\label{sec:obs_12CO(J=1-0)}
The details of the observations are described by \citet{Shimoikura_2019},  \citet{nakamura2019nobeyama} and \citet{Nguyen-Luong_2020}.
In brief, we carried out mapping observations in $^{12}$CO ($J = 1-0$) and $^{13}$CO ($J=1-0$)
toward the M17 region using the FOREST receiver \citep{Minamidani_2016}
installed on the 45 m telescope of the Nobeyama Radio Observatory. 
See \citet{nakamura2019nobeyama} for the actual mapping area. 
In this paper, we focus on the IRDC region.
The observations were done in an On-The-Fly (OTF) mode \citep{Sawada_2008} in the period from 2016 December to 2017 March. As the backends, we used a digital spectrometer based on an FX-type correlator, SAM45, 
that consists of 16 sets of 4096 channel array. The frequency resolution of all spectrometer arrays 
is set to 15.26 kHz, which corresponds to 0.05 km s$^{-1}$ at 110 GHz. 
The temperature scale was determined by the chopper-wheel method. The telescope pointing was checked every 1 hour by observing the SiO maser line. 
The pointing accuracy was better than $\sim 3\arcsec$ throughout the entire observation. 
The typical system noise temperature was in the range from 150 K to 200 K in the single sideband mode.
The main beam efficiency at 110 GHz was of $\eta_{mb}$ = 0.435.

In order to minimize the OTF scanning effects, the data with orthogonal scanning directions along the R.A. and Dec. axes were combined into a single map.
We adopted a spheroidal function with a width of 7\arcsec .5 as a gridding convolution function 
to derive the intensity at each grid point of the final cube data with a spatial grid size of 7.\arcsec 5, 
about a third of the beam size. The resultant effective angular resolution is about 22\arcsec  at 110 GHz, corresponding to $\sim 0.25$ pc at a distance of 2 kpc.

In this paper, we adopt the $^{12}$CO ($J=1-0$) peak temperature to derive excitation temperatures of CO molecules, and use the $^{13}$CO ($J=1-0$) emission to derive the column density at each pixel.
Se \citet{Nguyen-Luong_2020} for the CO peak intensity map of M17.
We assume the fractional abundance of $^{13}$CO relative to H$_2$ of $ 2 \times 10^{-6}$ \citep{Dickman_1978}, similar to in \citet{Shimoikura_2019} and \citet{Nguyen-Luong_2020}.

\subsection{$^{12}$CO $(J=3-2)$ observations}
\label{sec:obs_12CO(J=3-2)}
The $^{12}$CO ($J=3-2$) data were obtained with the James Clerk Maxwell Telescope(JCMT)
toward a much wider area than that of the $^{13}$CO ($J=1-0$) observations. The OTF mapping technique was employed. The observations were carried out from 2013 to 2015. All of the observations were 
carried out with the Heterodyne Array Receiver Program (HARP \citet{Buckle_2009}), 
consisting of 16 superconductor-insulator-superconductor detectors arranged 
in an array of 4 $\times$ 4 beams separated by 30\arcsec. At this frequency, the angular resolution of HARP is 14\arcsec and the main-beam efficiency of $\eta_{mb}$ = 0.61. 
The Auto-Correlation Spectral Imaging System (ACSIS; \citet{Buckle_2009}) was used as the backend with different frequency resolutions: $\sim$0.0305 MHz, $\sim$ 0.061 MHz, or $\sim$ 0.488 MHz.
We reduced the data with different frequency resolution separately. 
Data were reduced by the standard ORAC-DR pipeline on the Starlink package which contains various software for analyzing JCMT data
(\citet{Jenness_2015}; \citet{Berry_2007}; \citet{Currie_2008}). 
For more details on the data reduction, see \citet{Dempsey_2013} and \citet{Rigby_2016}. 
For this paper, we combined all of the three datasets and binned them to the coarsest spectral resolution of 0.488 MHz, that is of the COHR survey. Our final data for this paper have an angular resolution of 14\arcsec, a velocity resolution of 0.4 km s$^{-1}$, and are resampled onto the 7\arcsec $\times$ 7\arcsec grid. Detailed observation will be described in the following paper by \citet{Nguyen-Luong_2020}.

\section{Overall structure from $^{13}$CO $(J=1-0)$ emission}
\label{sec:overall}

\begin{figure*}[hbtp]
\begin{center}
\includegraphics[width=160mm]{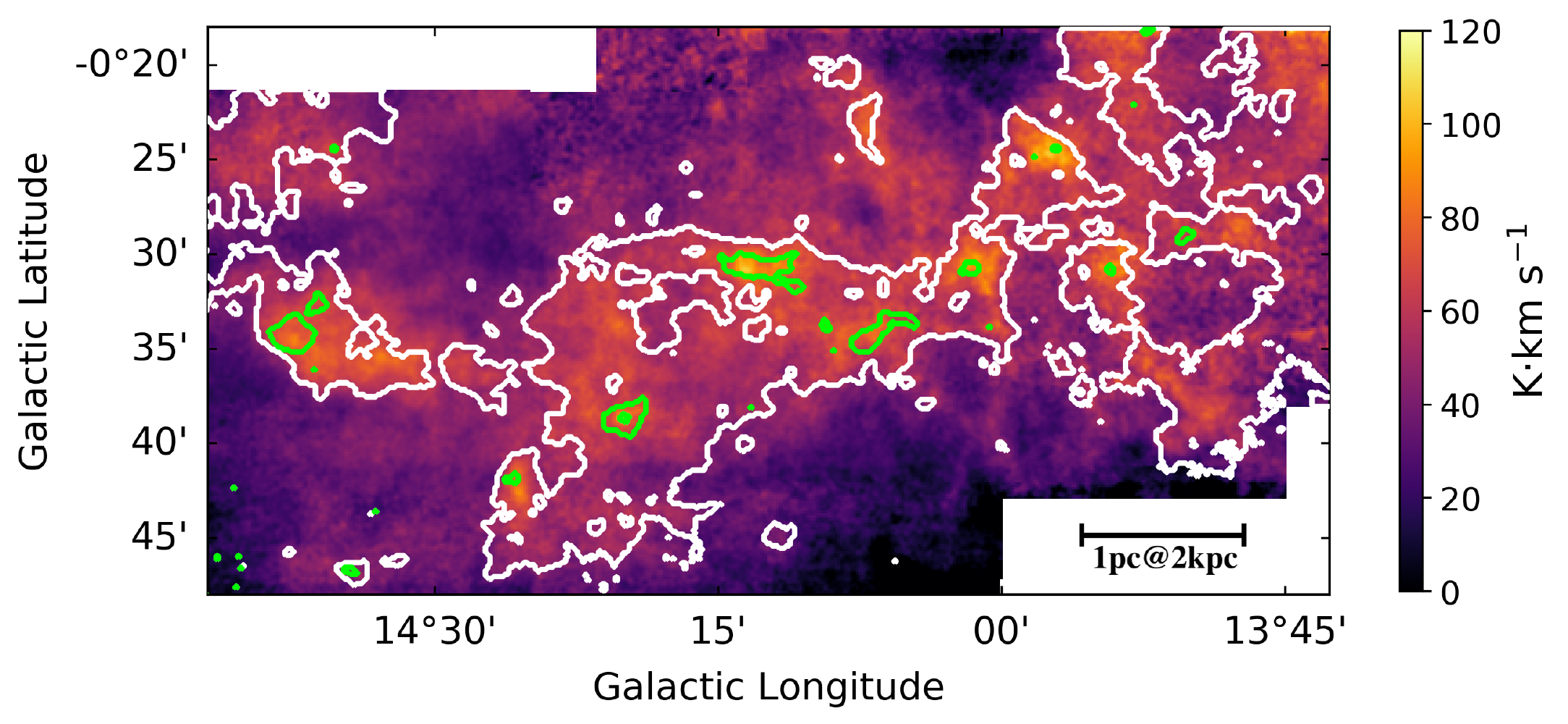}
\begin{flushleft}
\caption{The $^{13}$CO$(J=1-0)$ intensity map of M17 SWex, integrated over the velocity range from $-10$ km s$^{-1}$ to $55$ km s $^{-1}$. 
Contours of the H$_{2}$ column density map derived from the Herschel data are superimposed. Contours are drawn at 1.0$\times$ $10^{22}$cm$^{-2}$ by white, and 3.0$\times$10$^{22}$cm$^{-2}$ by green.}
\label{fig:all_mom0_and_cd}
\end{flushleft}
\end{center}
\end{figure*}

Figure \ref{fig:all_mom0_and_cd} shows the  $^{13}$CO $(J=1-0)$ velocity-integrated intensity map of M17 SWex. The contours of the H$_{2}$ column density map , which was derived  by SED fitting \citep{Sugitani_2019} of the Herschel archival data (160, 250, 350, and 500 $\mu$m) are superimposed.
The column density tends to be high in the areas with strong $^{13}$CO integrated intensities.
This indicates that the $^{13}$CO emission traces high density molecular gas in this region. 
The molecular gas is roughly distributed from southeast to northwest.

\begin{figure*}[hbtp]
\begin{center}
\includegraphics[width=110mm]{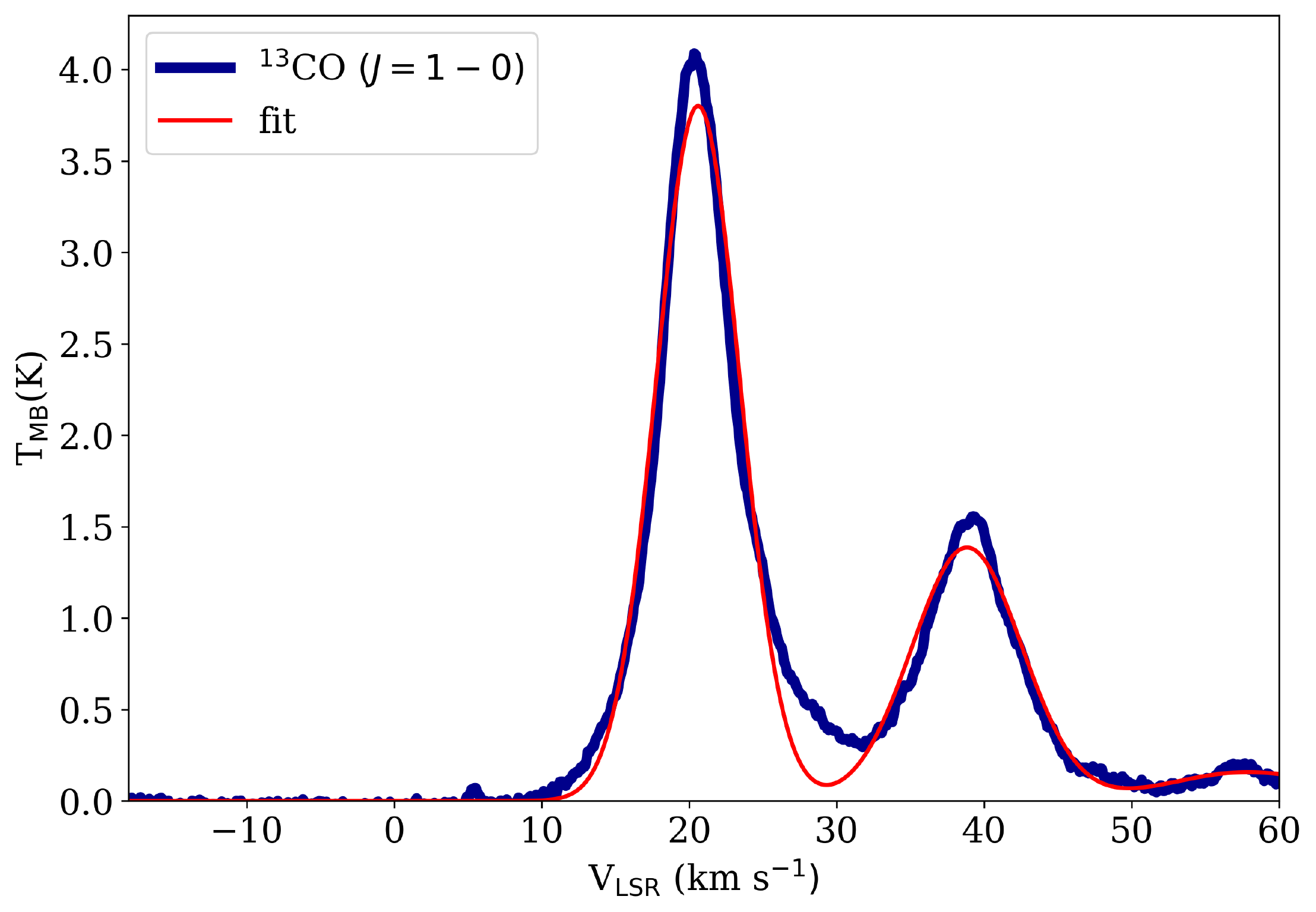}
\begin{flushleft}
\caption{The $^{13}$CO$(J=1-0)$ spectra averaged over the entire M17 SWex region (the deep blue thick line). 
The red line indicates the best-fitted three-Gaussian-component model, and the fitted parameters are listed in table\ref{tab:fitting_parameter}.
}
\label{fig:mom0-SWexregion-profile}
\end{flushleft}
\end{center}
\end{figure*}

Figure \ref{fig:mom0-SWexregion-profile} shows the $^{13}$CO $(J=1-0)$ spectrum averaged over M17 SWex. It has three major peaks over the velocity range from $-$18  km s$^{-1}$ to 60 km s$^{-1}$ (see also \citet{Nguyen-Luong_2020}). 
The red line shows the least-squares fit with three Gaussian components. 
These three component could correspond to the molecular gas which belong mainly to the Sagittarius, Scutum, and Norma arms, respectively \citep{nakamura2019nobeyama}. 
The strongest component of this region has a velocity of $\sim$21 km s$^{-1}$, and 
represents the main part of M17 SWex (\citet{Elmegreen_1979}; \citet{Nguyen-Luong_2020}).
We did not fit the small spike around 6 km s$^{-1}$.

\begin{figure*}[hbtp]
\begin{center}
\includegraphics[width=170mm]{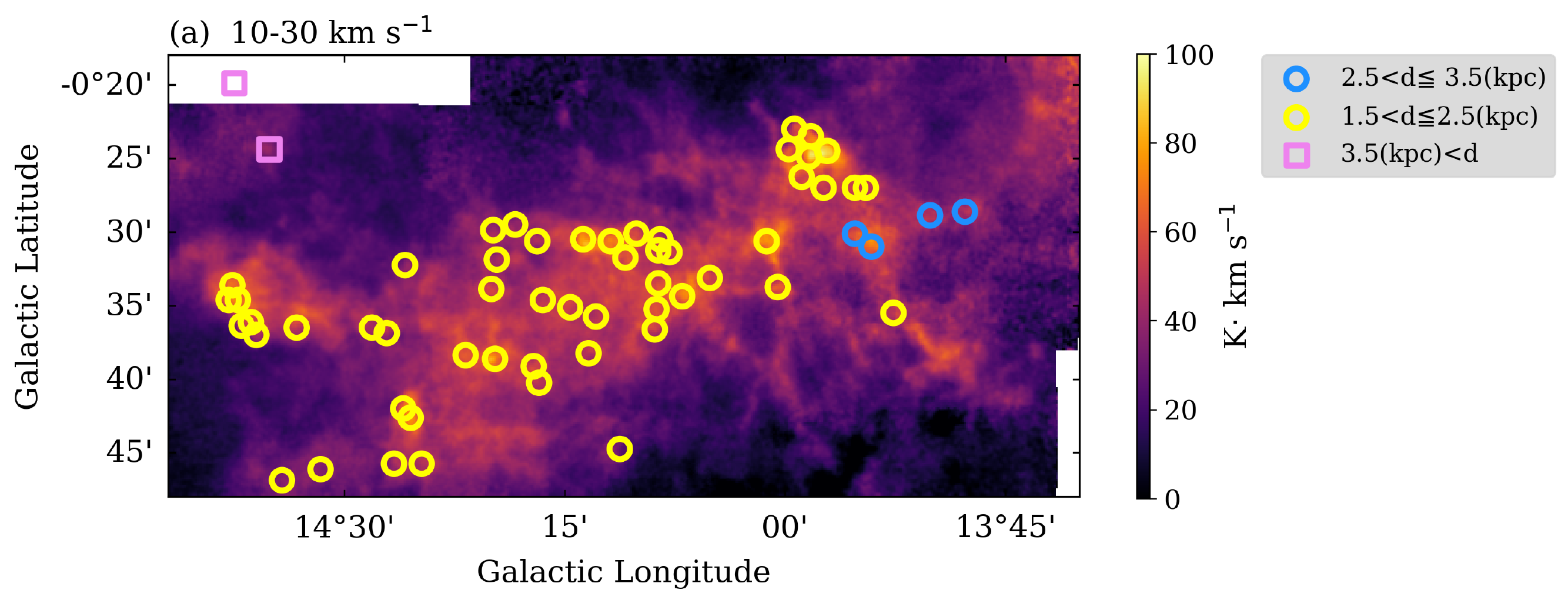}
\includegraphics[width=170mm]{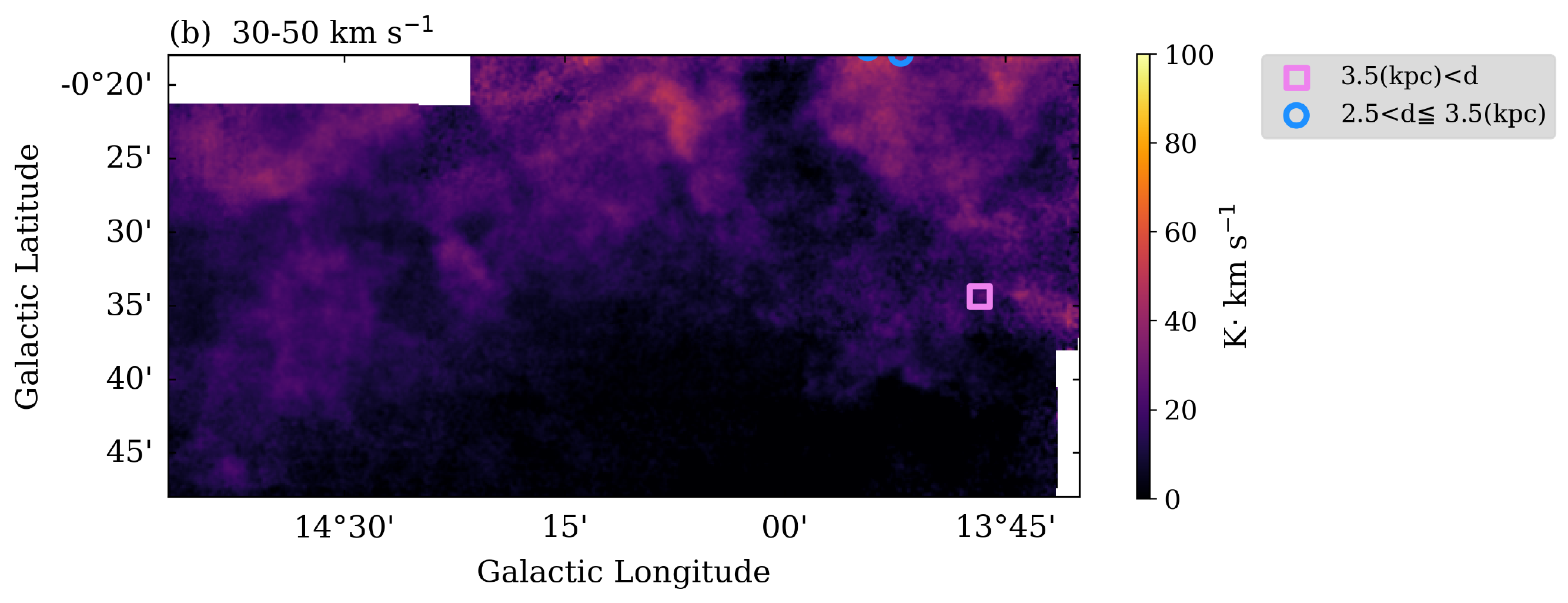}
\begin{flushleft}
\caption{(a) The $^{13}$CO($J=1-0$) intensity map integrated from 10 km s$^{-1}$ to
30 km s$^{-1}$. (b) same as panel (a) but for the velocity range of 
30 km s$^{-1}$ $<V_{\rm LSR}<$50 km s$^{-1}$. In both panels, clumps cataloged by \citet{Urquhart2018} are presented, which have the mean velocity within the velocity range used for the integration. Shapes and colors of each mark represent the distances.}
\label{fig:mom0_SWexregion}
\end{flushleft}
\end{center}
\end{figure*}

Figure \ref{fig:mom0_SWexregion} shows the $^{13}$CO $(J=1-0)$ intensity maps of M17 SWex, integrated over the velocity ranges from 10 km s$^{-1}$ to 30 km s$^{-1}$, and from 30 km s$^{-1}$ to 50 km s$^{-1}$ (see also figure 18 in \citet{Nguyen-Luong_2020}).
The bulk of the M17 SWex emission is seen in the velocity range 10 $-$ 30 km s$^{-1}$.
The emission in the velocity range 30 $-$ 50 km s$^{-1}$ is stronger toward the
galactic plane (the upper area (around $b=0$) in each panel). 

For comparison, we also plotted clumps identified with ATLASGAL, an unbiased 870 micron submillimetre survey of the inner galactic plane and provides a large and systematic inventory of all massive  ($>$1000 $M_{\odot}$), dense clumps in the Galaxy \citep{Schuller_2009}. 
\citet{Urquhart2018} derived the detailed properties (velocities, distances, luminosities and masses) and spatial distribution of a complete sample of $\sim$ 8000 dense clumps detected with the ATLASGAL survey. The distances of the clumps are determined by using various datasets such as maser parallax and spectroscopic data of HI and molecular lines.  The uncertainty on the distance estimation may 
be $\sim$ 0.3 kpc or more. Therefore, we expect that we can distinguish 
between the structures associated with different arms. 
In the following, we adopt the distances listed in \citet{Urquhart2018} (Table 2, Column 7) toward the clumps located in this area.
The ATLASGAL clumps at $\sim 2$ kpc are associated mainly with the cloud structure in the velocity range from 10 km s$^{-1}$ to 30 km s$^{-1}$. In the upper edge of figure 3 (b), there are two clumps $\sim$ 3 kpc. As shown in figure \ref{fig:12CO(J=3-2)map} (b) (appendix \ref{app:Atlasgal_clump}), the majority of the ATLASGAL clumps located near the Galactic plane have distances of $\sim$ 3 kpc, and are closely associated with $^{13}$CO structures within the 30 $-$ 50 km s$^{-1}$ velocity range toward the M17 SWex cloud structure.
Therefore, we infer that most of the $^{13}$CO emission in this velocity range is likely to originate from the Scutum arm in the background.
However, as is discussed later (section \ref{sec:ccc}), there may be some 30 $-$ 50 km s$^{-1}$ cloud structures located at the same distances as M17 SWex ($\sim$ 2kpc). 


\section{Cloud Identification with SCIMES}
 \label{sec:result}
 
\subsection{Structures identified with SCIMES}
 We identify cloud structure in M17 SWex by applying the  SCIMES (Spectral Clustering for Interstellar Molecular Emission Segmentation, \citet{Colombo_2015} ) to the $^{13}$CO $(J=1-0)$  data cube. 
 SCIMES is an algorithm based on graph theory and cluster analysis. SCIMES identifies relevant molecular gas structures within dendrograms \citep{Rosolowsky_2008} of emission using the spectral clustering paradigm.

First, we identify the cloud structures with dendrograms using the following three parameters 
{\tt  min$\_$value}=10 $\sigma_{\rm rms}$, 
{\tt min$\_$delta}=3 $\sigma_{\rm rms}$, and 
{\tt min$\_$npix}=30, where  $\sigma_{\rm rms}$=0.38 K is the average rms noise level of the $^{13}$CO $(J=1-0)$ data.
The first parameter, {\tt  min$\_$value}, represents the minimum value of the intensity used for analysis. Above this minimum, structures are identified. The second parameter, {\tt min$\_$delta}, is the minimum step of intensity required for a structure to be identified. {\tt min$\_$delta} defines a minimum significance for structures. The third parameter, {\tt min$\_$npix}, is the minimum number of pixels in the position-position-velocity space that structure must contain. 
Then, we applied SCIMES for the hierarchical structures specified by dendrograms.
In total, we identified 118 individual structures. Hereafter, we call structures identified by SCIMES "clouds".

\begin{figure*}[hbtp]
\begin{center}
\includegraphics[width=170mm]{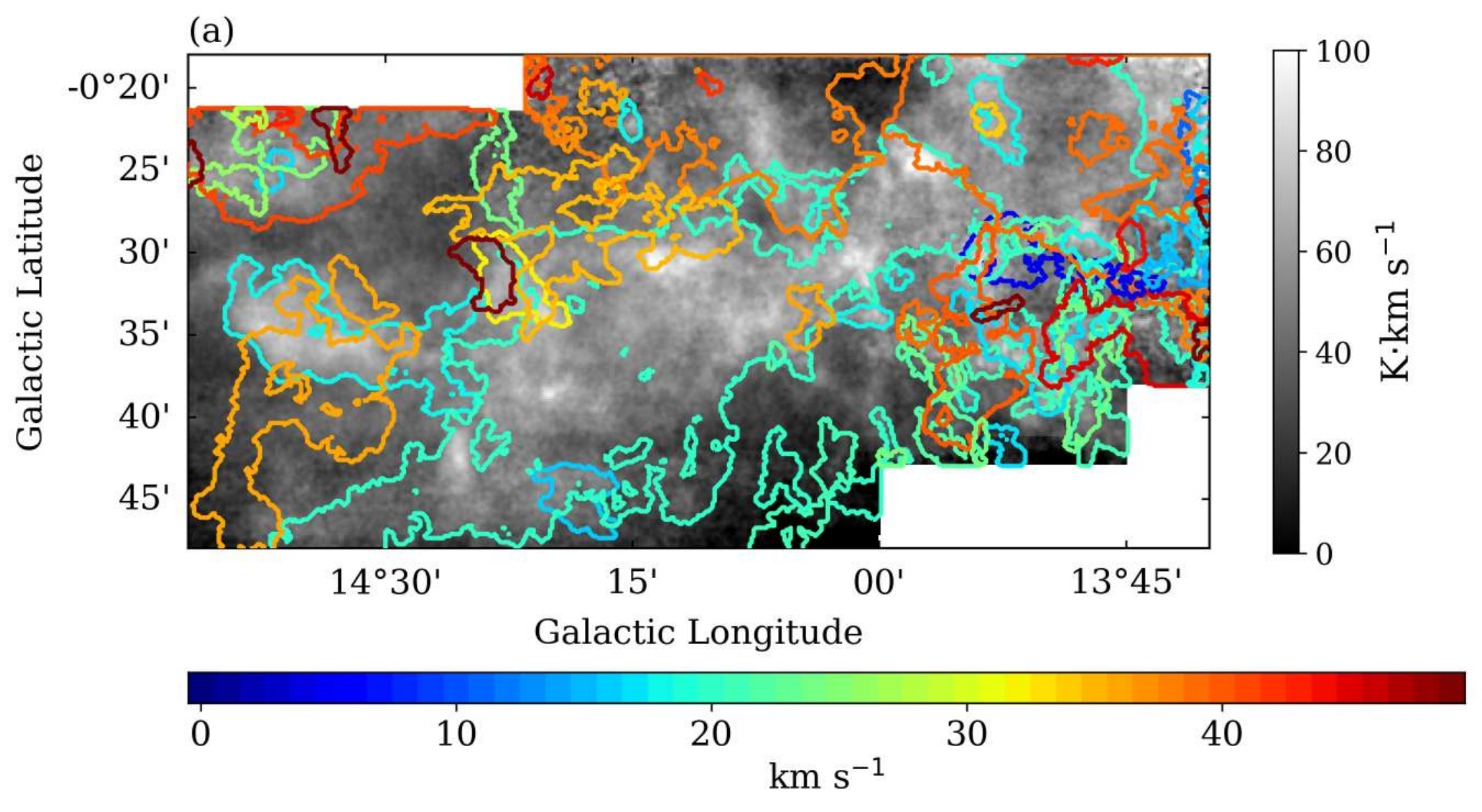}
\includegraphics[width=170mm]{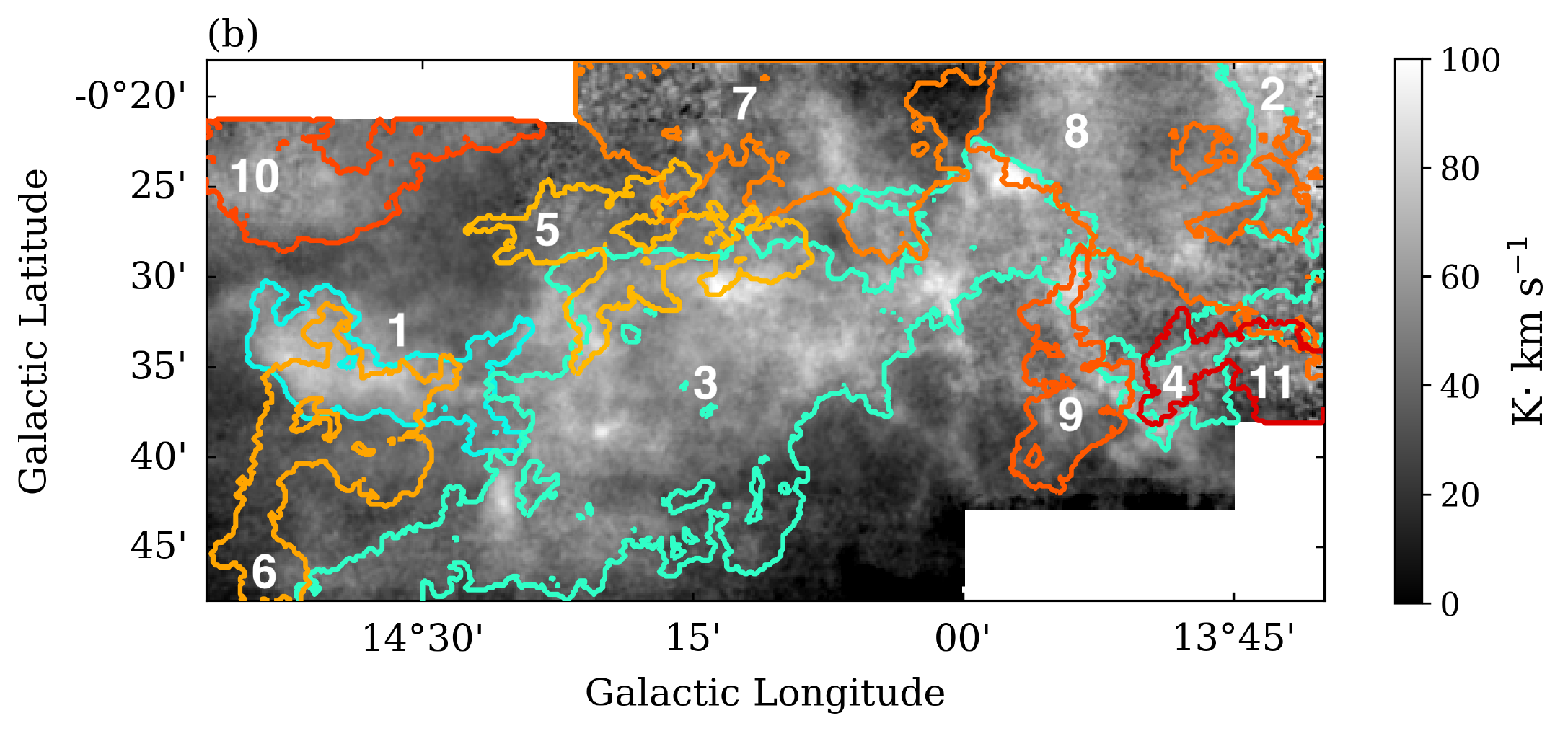}
\begin{flushleft}
\caption{Structures indentified by SCIMES. Contours represent outline of every structure orthogonally projected on the plane of the sky. Each contour is colored depending on its mean velocity of the structure, according to the colorbar at the bottom of this panel. The background image is the integrated intensity map of the $^{13}$CO$(J=1-0)$. The velocity range used for the integration is $-$18 km s$^{-1}$ $<V_{\rm LSR}<$60 km s$^{-1}$. (a) All clouds identified by SCIMES are shown as countoured. (b) Only clouds with radii $>$ 1.7${}^{\prime}$ (equivalent to 1 pc at a distance of 2 kpc) are shown. The labels denote the ID in table \ref{tab:large-size structure}.
}
\label{fig:scimes}
\end{flushleft}
\end{center}
\end{figure*}

\begin{figure*}[hbtp]
\begin{center}
\includegraphics[width=170mm]{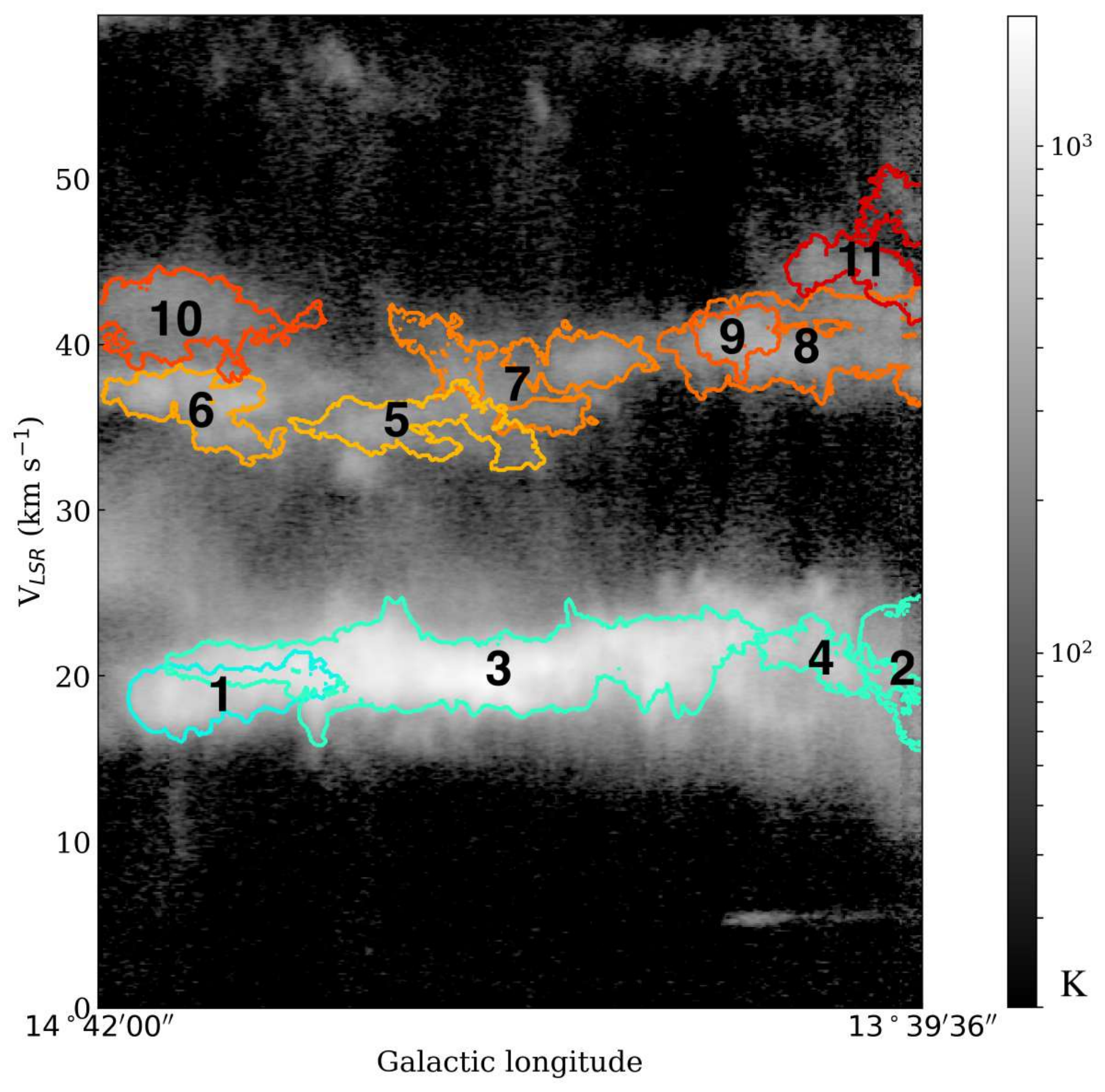}
\begin{flushleft}
\caption{Structures identified by SCIMES with radii greater than $1.7^{\prime}$ in velocity-galactic latitude diagram. Each contour represents the identified structures projected on the velocity-galactic plane . Colors of the contours are the same as in figure \ref{fig:scimes}. The background image is the $^{13}$CO$(J=1-0)$ data integrated in galactic latitude direction.}
\label{fig:scimes_p-v}
\end{flushleft}
\end{center}
\end{figure*}

Figure \ref{fig:scimes}(a) shows the spatial distribution of clouds identified by SCIMES.
The color represents the mean velocities of the clouds. 
In figure \ref{fig:scimes}(b), we show only the clouds whose radii are greater than 
1.7${}^{\prime}$ (equivalent to 1 pc at a distance of 2 kpc).
Some properties of these large clouds are listed in table \ref{tab:large-size structure}.
Figure \ref{fig:scimes_p-v} represents the distribution of clouds with radii greater than $1.7^{\prime}$ in position-velocity (P-V) diagram which is integrated in the galactic latitude direction.

It is worth noting that the cloud identification depends on the dendrograms parameters. 
For example, if we adopt a smaller value of {\tt  min$\_$value},  the No. 5 and No. 6 clouds become a single structures. 
We discuss how the structure identification changes with other parameters in the appendix \ref{app:5sigma_scimes}. In this paper, 
we adopt a relatively-high {\tt  min$\_$value} of 10$\sigma$ to inspect relatively-high-density structure embedded in less dense structure.

Figure \ref{fig:scimes_p-v}  shows that mean velocities of most of the clouds are distributed around 20 km s$^{-1}$ or 40 km s$^{-1}$.
The main component in M17 SWex is shown in light green (around 20 km s$^{-1}$), resembling a "flying dragon" in figure \ref{fig:scimes} (No. 3 in table \ref{tab:large-size structure}).
The cyan colored structure (No. 1 in table \ref{tab:large-size structure}) touches this main component and has a mean velocity around 20 km s$^{-1}$ too.
These two clouds (No. 1 and No. 3) also become a single structure if we adopt a lower {\tt  min$\_$value}. Table \ref{tab:focusing_structure} shows the properties of these clouds.

\subsection{Comparison with the magnetic field directions}
\label{subsec:magnetic_field-result}

\begin{figure*}[hbtp]
\includegraphics[width=170mm]{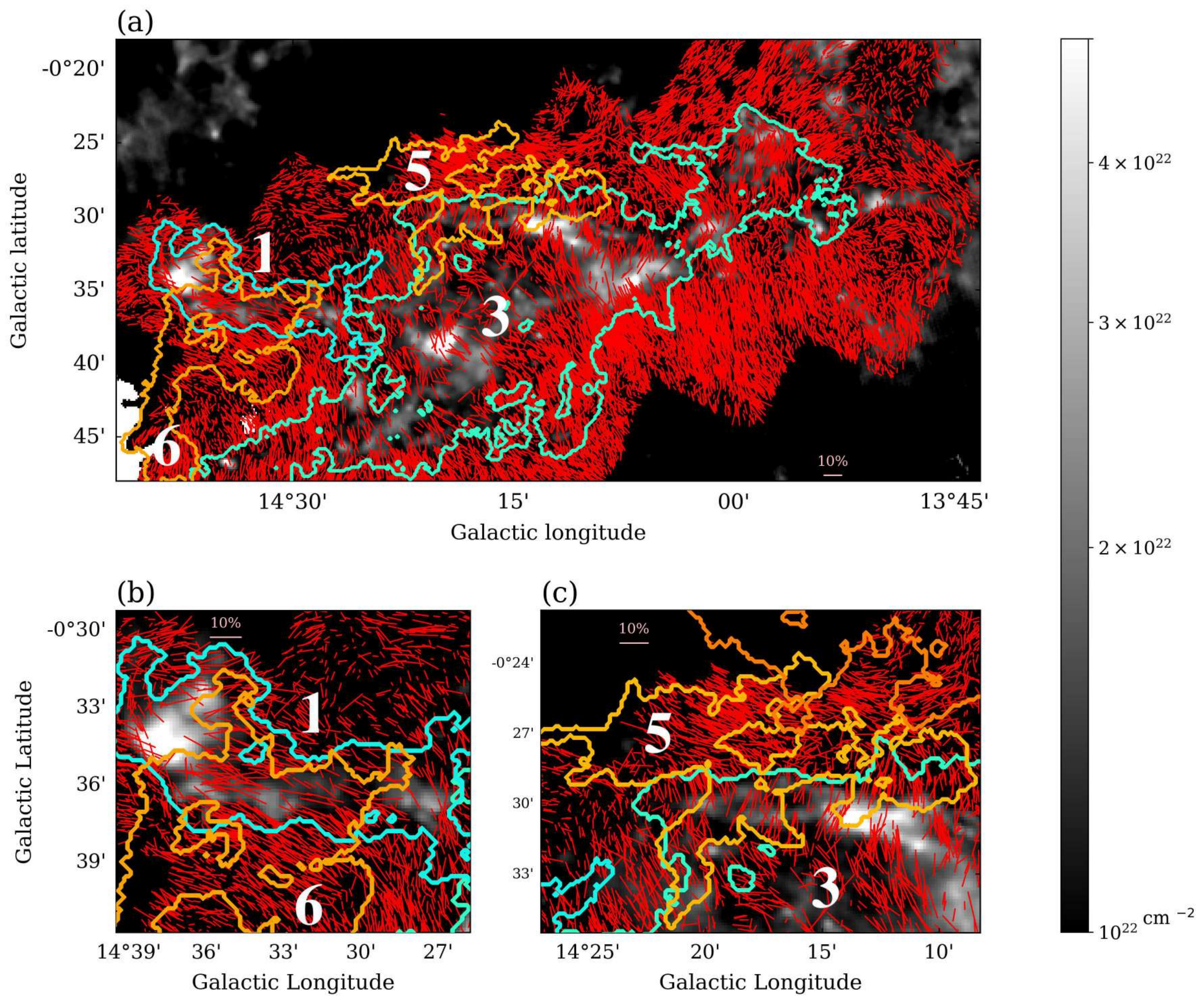}
\begin{flushleft}
\caption{Near-IR polarization H-band vector maps \citep{Sugitani_2019} superposed on the H$_{2}$ column density map derived from the Herschel data. The length of each vector is in proportion to it’s polarization degree and a vector of 10\% polarization is shown. Colored contours represent the No 1,3,5 and 6 cloud structures (see table \ref{tab:focusing_structure}). (a) Over all view of the M17 SWex. (b),(c) Enlarged view of interaction areas of the No. 1/6 and 3/5 clouds.   }
\label{fig:magnetic_field}
\end{flushleft}
\end{figure*}

\citet{Sugitani_2019} performed near infrared polarization observations toward the M17 SWex area and revealed that the global magnetic fields are roughly perpendicular to the galactic plane and the dense filamentary structures. Figure \ref{fig:magnetic_field} shows near-IR polarization H-band vector maps \citep{Sugitani_2019} superposed on the H$_{2}$ column density map.
The magnetic field is globally perpendicular to structure No. 3 extending from southeast to northwest. However, the magnetic fields suddenly change their directions at the intersection areas of the No. 1/6 and 3/5 clouds, and the field direction becomes preferentially parallel to the galactic plane.
In other words, the magnetic field vectors appear to change their directions at the interface of the clouds.

\subsection{Intermediate velocity components}
\label{sec:intermediate}
we present in figure \ref{fig:p-v_12CO} (a) the $^{12}$CO $(J=1-0)$ and $^{13}$CO $(J=1-0)$ intensity maps integrated in the range 
24.6 km s$^{-1}$ $<V_{\rm LSR}<32.5$ km s$^{-1}$ which corresponds to the intermediate velocity range between No. 1/3 and 5/6 clouds.
For comparison, we show the boundaries of the No. 1, 3, 5 and 6 clouds by the contours. We focus on clouds Nos. 1, 3 and Nos. 5, 6 respectively in table \ref{tab:large-size structure}, because clouds Nos. 1, 3 are main components of M17 SWex, and  Nos. 5, 6 are overlapped or close to these components.

In figure \ref{fig:p-v_12CO} (b), we show the P-V diagram along the white solid 
line indicated in panel (a). Figure \ref{fig:p-v_12CO} (c),(d) show the mean P-V (galactic latitude-velocity) diagrams within the range indicated by the respective white dashed lines in panel (a).
The $^{12}$CO emission is intense around the 20 km s$^{-1}$ (No. 1 and No. 3)
and 35 km s $^{-1}$ (No. 5 and No. 6) components, and the faint emission is mainly distributed
between these two velocities. In contrast, outside the two components ($>$ 40 km s$^{-1}$ and
$<$ 15 km s$^{-1}$), such faint emission is rare to find. 

\section{Discussion and Conclusions}
\label{sec:ccc}

As mentioned above, clouds with different velocities appear to overlap along the line of sight.
Figure \ref{fig:mixed_intensity-map} shows $^{13}$CO $(J=1-0)$  contour plot of intensity integrated from 10 km s$^{-1}$ to 30 km s$^{-1}$ (green contours) and 30 km s$^{-1}$ to 50 km s$^{-1}$ (red map) (see also \citet{Sugitani_2019}). 
From the cloud morphology, \citet{Sugitani_2019} pointed out that a couple of clouds with 30$-$50 km $s^{-1}$ 
 appears to physically interact with those with 
10$-$30 km $^{-1}$. 
If this is the case, the 30 $-$ 50 km s$^{-1}$ component should be located at the same distance as the 10$-$30 km $^{-1}$ component, 2 kpc.

Here, we discuss three possible scenarios that appear to be consistent with observational properties: (i) overlapped clouds with different distances, (ii) oscillation of a larger cloud, and (iii) cloud-cloud collision.

\subsection{Case (i): a chance coincidence of clouds with different distances}
As mentioned in section \ref{sec:overall}, 
some molecular clouds with $>$ 35 km s$^{-1}$ are likely to belong mainly 
to the distant Scutum or Norma spiral arms and simply overlap with the main component with 
$\sim$20 km s$^{-1}$ along the line-of-sight. This may be the case for clouds 5 and 6. In subsection \ref{sec:intermediate}, we presented the intermediate velocity components in the P-V diagram. There is a possibility that this emission comes from the inter-arm region between the Sagittarius arm and Scutum arm. 
The bending magnetic fields at the intersection of the clouds may be consistent with the observed polarization pattern if the orientation of the magnetic fields associated with the individual clouds are different.  For example,
if the magnetic field orientations of the No. 5/6 clouds are preferentially parallel to the Galactic plane, and the magnetic fields of the No. 1/3 clouds are perpendicular to the Galactic plane, the observed polarization pattern would be achieved.


\subsection{Case (ii): a larger cloud oscillation}

One of the possible scenarios which can explain the existence of CO structures with a peak velocity of $\sim$35km/s (No. 5/6 clouds) and the spatial distribution of these structures with $\sim$20km/s structures is that these structures are denser parts created by the oscillation of a single larger cloud. 
Such oscillation of clouds has been discussed for a Bok globule, B 68 \citep{redman_2006}. 
If a cloud is oscillating, the clouds identified should be gravitationally bound as a whole. Here we discuss the dynamical state of the two clouds with $V_{\rm LSR}$ $\sim$ 20 km s$^{-1}$ and $\sim 35$ km s$^{-1}$ using a simple analysis. 

To verify the gravitational state of these two cloud complexes, we estimated the mass required to gravitationally bind the structures using the following equation
\begin{equation}
\label{eq:res_mass}
  M_{\rm res}=\frac{\Delta V_{3D}^2 R}{G} ~, 
\end{equation}
where $G$ is the gravitational constant, $R$ is the separation of the two cloud complexes, $\Delta V_{3D}$ is the 3D relative velocity. 
A cloud having a mass greater than $M_{\rm res}$
is expected to be gravitationally bound.
We assume the separation of the two cloud complexes is $\sim$ 5.1 pc which is the distance between the mean positions of the No. 3 and  No. 5 clouds. The 3D relative velocity is calculated as $\Delta V_{3D}=\sqrt{3}\Delta V_{1D}$. Inserting in Equation (\ref{eq:res_mass}), $\Delta V_{3D} \sim$ $\sqrt{3}\times$ 15 km s$^{-1}$and $R\sim$5.1 pc, we obtain $M_{\rm res}\sim~8.0\times10^5 M_{\odot}$. This mass is larger than the actual cloud mass $\sim$ 4.5 $\times 10^{4} M_{\odot}$ by a factor of $\sim$ 10.
Therefore, this system contained clouds identified
does not appear to be gravitationally bound. In other words, the oscillation of a larger cloud appears to be unlikely.
However, because the physical values used in the above analysis have large uncertainties from, e.g., the inclination angle of the clouds respect to the line of sight, and the distance of M17 SWex, it is difficult to rule out this possibility completely.

\subsection{Case (iii): cloud-cloud collision}
\label{subsec:collision_scenario}

Here, we discuss the possible collision of $\sim$ 20 km s$^{-1}$ clouds and  $\sim$ 35 km s$^{-1}$ clouds. 
We provide two primary pieces of evidence supporting the cloud-cloud collision scenario. One is a bridge feature in the P-V diagram which often appears in the early stages of cloud-cloud collisions \citep{Takahira_2014,Haworth_2015}.
The other is bent magnetic field structures which appear to consistent with numerical simulation
results of the colliding magnetized clouds \citep{Wu_2017}.

The extended emission with the intermediate velocities shown in figure \ref{fig:p-v_12CO} implies that the two structures may be dynamically interacting. 
\citet{Haworth_2015} demonstrated that the turbulent motion in the compressed layer 
of a cloud-cloud collision can be observed in a P-V diagram as "broad bridge features" 
connecting the two clouds. They produced synthetic P-V diagrams from a range of different simulations of molecular clouds, including cloud-cloud collisions and isolated clouds. They found that "broad bridge features" appeared in their cloud-cloud collision models, but did not appear in any of the simulations of isolated clouds (see also \citet{Takahira_2014}).
Thus, the No. 5 and No. 6 clouds might lie at the same distance 2kpc as No. 1 and No. 3, and these four clouds might be colliding with each other.

Appendix \ref{app:non-bridge} shows mean P-V (galactic latitude-velocity) diagrams measured in regions where No. 7 and 8 clouds overlap with No.3. In these diagrams, emission at the intermediate velocities is very weak, and there are no bridge features detected. So, No.7 and No.8 clouds may be unrelated to the cloud-cloud collision.

Additionally, No.7 and No.8 clouds are closer to b=0 degree than No. 1/3 clouds. Since the solar system is located several hundred pc above the midplane of our Galaxy, M17 SWex should appear displaced toward more negative or positive latitudes, while distant clouds appear around b = 0 degree due to perspective effects. Hence, No.7 and No.8 clouds may be more distant clouds. While, No 5/6 clouds are actually at the similar latitudes to No 1/3 clouds, this fact is consistent with the idea that No 5/6 clouds are located at the same distance as No 1/3 clouds.

We also found noteworthy features of the magnetic field orientation in some possible interacting parts. As discussed in section \ref{subsec:magnetic_field-result}, the magnetic field orientation changes abruptly at the intersection areas of the No. 1/6 and 3/5 clouds. \citet{Wu_2017} indicate that when two clouds collide, magnetic fields become oriented preferentially parallel along the boundary of collision. The observed abrupt changes of the magnetic field seem qualitatively consistent with the cloud-cloud collision scenario.

Two clouds, No. 5 and No. 6 lie adjacent to each other and have similar velocities.
Further inspection of the channel maps (see appendix \ref{app:channelmap}) reveals weak emission connecting these two clouds
Therefore, we can interpret that the No. 5 and No. 6 clouds may be connected and belong to a larger structure, but two dense parts within one structure may have been identified independently with SCIMES. 
The same goes for the No. 1 and No. 3 clouds.
The masses of the No. 1 and No. 3 clouds are estimated to be 7.5$\times 10^{3} M_{\odot}$, and 3.7$\times 10^{4} M_{\odot}$, respectively (see table \ref{tab:focusing_structure}), while 
the masses of the No. 5 and No. 6 clouds are estimated to be 1.3$\times 10^{3} M_{\odot}$, and 1.7$\times 10^{3} M_{\odot}$, respectively. If the No. 1/No. 3 clouds and the No. 5/No. 6 clouds are single structures, the two large complexes, have masses greater than  4.5$\times 10^{4} M_{\odot}$ and 3.0$\times 10^{3} M_{\odot}$, respectively.
In summary, we propose that a $\sim$20 km s$^{-1}$ cloud complex with $\gtrsim$ 4.5 $\times 10^{4} M_{\odot}$ and a $\sim$35 km s$^{-1}$ cloud complex with $\gtrsim$ 3.0 $\times 10^{3} M_{\odot}$ may be colliding with a relative speed of $\sim 15$ km s$^{-1}$ are colliding for case (iii).

In summary, we think that the most plausible scenario is the case (iii), the cloud-cloud collision, although we just presented only circumstantial evidence such as the spatial configuration of the clouds, a bridge emission, and bending magnetic field at the possible interacting areas. Further concrete evidence would be needed to prove the cloud-cloud collision scenario completely.

\section{Summary}
\label{sec:summary}

We analyzed the cloud structure of the M17 SWex in $^{13}$CO ($J=1-0$).
Our main results are summarized as follows:
\begin{itemize}
\item[1.] Our $^{13}$CO integrated intensity distribution in M17 SWex closely follows the dense part traced by the Hershel column density map. 
\item[2.] By applying SCIMES to the $^{13}$CO data cube, we identified 118 clouds in M17 SWex.
\item[3.] We select 6 large ($>$1.7${}^{\prime}$) clouds identified by SCIMES. 
At least two large clouds with $\sim 35$ km s$^{-1}$ seem to be located close to the main clouds at $\sim 20$ km s$^{-1}$ in the galactic longitude$-$galactic latitude plane.

\item[4.] We discussed three possibilities that appear to be consistent with the observed features. The first scenario is that clouds with different distances are overlapped along the line of sight. The second scenario is that a large cloud is oscillating.  The third scenario is that clouds located at the same distances are colliding.
\item[5.]Judging from (1) the existence of the bridge feature in the P-V map, and (2) the distortion of the magnetic field orientation at the intersections, we think that cloud-cloud collision is the most plausible scenario. However, it is very difficult to rule out other scenarios. Further concrete evidence would be needed to prove the cloud-cloud collision scenario.




\end{itemize}

\section{Acknowledgements}

This work was partly supported by JSPS KAKENHI Grant Numbers JP24540233, JP16H05730 and JP17H01118.

This work was carried out as one of the large projects of the Nobeyama Radio Observatory (NRO), which is a branch of the National Astronomical Observatory of Japan, National Institute of Natural Sciences. 

We thank the NRO staff for both operating the 45 m and helping us with the data reduction. 

B.W and P.S. were partly supported by a Grant-in-Aid for Scientific Research (KAKENHI Number 18H01259) of Japan Society for the Promotion of Science (JSPS).

GJW expresses his grateful thanks to The Leverhulme Trust for an Emeritus Fellowship.

\clearpage


\appendix

\section{Integrated intensity maps of $^{12}$CO$(J=3-2)$ with ATLASGAL clumps}
\label{app:Atlasgal_clump}
\vspace{5pt}
Figure \ref{fig:12CO(J=3-2)map} shows the integrated intensity maps of the $^{12}$CO$(J=3-2)$ emission line. The velocity ranges used for the integration are (a) 10 km s$^{-1}$ $<V_{\rm LSR}<30$ km s$^{-1}$ and (b) 30 km s$^{-1}$ $<V_{\rm LSR}<50$ km s$^{-1}$. In both panels, we plot clumps cataloged by \citet{Urquhart2018}.

\section{Identified clouds with SCIMES for different parameters of dendrograms}
\label{app:5sigma_scimes}
\vspace{5pt}
Figure \ref{fig:scimes-5sigma} shows identified clouds with SCIMES , when using the three parameters of {\tt  min$\_$value}=5 $\sigma_{\rm rms}$, {\tt min$\_$delta}=3 $\sigma_{\rm rms}$, and 
{\tt min$\_$npix}=30.

\section{Channel maps of $^{13}$CO $(J=1-0)$ of the M17 SWex}
\label{app:channelmap}
\vspace{5pt}
Figure \ref{fig:channelmap} shows the $^{13}$CO $(J=1-0)$ channel maps integrated over 1 km s$^{-1}$ in the range 10$-$50 km s$^{-1}$. For comparison, in each map, we show the boundaries of No. 1, 3, 5 and 6 clouds in the corresponding velocity ranges 
(see tables \ref{tab:large-size structure} and \ref{tab:focusing_structure}) by contours.

\section{P-V diagram including No.3 and No. 7/8 clouds}
\label{app:non-bridge}
Figure \ref{fig:non-bridge} (a) shows the $^{12}$CO$(J=1-0)$ integrated intensity map integrated over the range 24.6$-$36.4 km s$^{-1}$.  In figures \ref{fig:non-bridge} (b) and (c), the upper panels show the P-V diagrams of the $^{12}$CO$(J=1-0)$ emission line taken within the range white broken lines indicate in panel (a). The bottom panel shows the mean spectra within the range between the two white broken lines in the upper panel. Velocity ranges of the identified clouds are shaded.

\section{Comparison with the N$_2$H$^+$ cores}
\label{app:core}Recently, \citet{Shimoikura_2019} identified 46 dense cores and clumps in M17 SWex by using the N$_2$H$^+$
($J=1-0$) line, and constrain their evolutionary times by comparing with the evoluytionary stages of YSOs classified by \citet{Povich_2016}.
Figure \ref{fig:core_column} shows the distribution of the dense  cores and clumps identified by \citet{Shimoikura_2019} with the stages classified. The background image shows the H$_{2}$ column density distribution derived from the Herschel data. 
The dense cores and clumps are mainly distributed along the two large filamentary structures seen in the H$_{2}$ column density map.

Here, we discuss whether a possible cloud-cloud collision  triggered the formation of dense structures such as cores and clumps in this region.
Here, we define the cloud-cloud collision timescale as the time required for clouds with a relative velocity $V_{\rm rel}$ to cross each other. We adopt the typical diameter of the No 1/3 clouds
($\sim$1.5-12 pc) and the relative velocity of 15 km s$^{-1}$  (see table \ref{tab:large-size structure}). 
The cloud-cloud collision timescale is roughly estimated to be $\sim$0.05 $-$ 0.4 Myr.
If the evolutionary times of the dense cores and clumps are shorter than the crossing time, there is a possibility that the formation of cores and clumps may have been triggered by the collision.

The collision timescale of $< 10^6$ yr seems to be shorter than ages of stage $\rm{II/III}$ stars \citep{Schulz_2012}. Thus, it is difficult to form protostellar cores associated with stage $\rm{II/III}$ YSOs by the cloud-cloud-collision we discussed here. However, several cores with stage $\rm{0/I}$ YSOs and cores with no YSOs, which seem to be distributed mainly in the overlapped areas of Nos. 1, 3, 5, and 6 clouds might be formed by the cloud-cloud collision.

\bibliography{reference}

\newpage

\begin{table*}
\tbl{Gaussian parameters best fitting the averaged spectra in figure \ref{fig:mom0-SWexregion-profile}}{
\begin{tabular}{cccc}
\toprule
{} &     $V_{\rm LSR}$ &                $\rm T_{\rm MB}$ &     $\sigma_{v}$ \\
{} &     (km s$^{-1}$) &              (K) &    (km s$^{-1}$) \\
\midrule
Gaussian 1 &  $20.61 \pm 0.02$ &  $3.80 \pm 0.02$ &  $2.84 \pm 0.02$ \\
Gaussian 2 &  $38.84 \pm 0.06$ &  $1.39 \pm 0.02$ &  $3.73 \pm 0.07$ \\
Gaussian 3 &  $57.91 \pm 2.15$ &  $0.16 \pm 0.02$ &  $5.40 \pm 1.86$ \\
\bottomrule
\end{tabular}
}
\vskip10pt
\begin{tabnote}
\end{tabnote}
\label{tab:fitting_parameter}
\end{table*}

\begin{table*}
\tbl{Clouds identified with SCIMES whose radii are more than 1.7${}^{\prime}$ (=102.0 ${}^{\prime\prime}$, 1 pc at 2 kpc)}{
\begin{tabular}{cccccccc}
\toprule
No $^{a}$& ${v_{\rm cen}}^{b}$ &                                  $l$ $^{c}$ &                                  $b$ $^{d}$ & R $^{e}$ & R$\_$major $^{f}$ & R$\_$minor $^{g}$ & boundary $^{h}$ \\
{} & \multicolumn{3}{l}{(km s$^{-1}$)} &      (arcsec) &               (arcsec) & \multicolumn{2}{l}{(arcsec)} \\
\midrule
1  &               18.79 &  $14^\circ27{}^\prime00{}^{\prime\prime}$ &  $-0^\circ36{}^\prime36{}^{\prime\prime}$ &        137.91 &                 233.71 &                  81.38 &                 \\
2  &               20.50 &  $13^\circ40{}^\prime12{}^{\prime\prime}$ &  $-0^\circ20{}^\prime24{}^{\prime\prime}$ &        102.48 &                 138.72 &                  75.71 &               Y \\
3  &               20.51 &  $14^\circ13{}^\prime12{}^{\prime\prime}$ &  $-0^\circ31{}^\prime12{}^{\prime\prime}$ &        342.90 &                 602.85 &                 195.04 &               Y \\
4  &               20.59 &  $13^\circ40{}^\prime12{}^{\prime\prime}$ &  $-0^\circ31{}^\prime48{}^{\prime\prime}$ &        130.63 &                 227.00 &                  75.17 &               Y \\
5  &               35.03 &  $14^\circ19{}^\prime12{}^{\prime\prime}$ &  $-0^\circ30{}^\prime00{}^{\prime\prime}$ &        170.89 &                 258.63 &                 112.91 &                 \\
6  &               36.21 &  $14^\circ38{}^\prime24{}^{\prime\prime}$ &  $-0^\circ46{}^\prime12{}^{\prime\prime}$ &        188.21 &                 265.48 &                 133.43 &                 \\
7  &               38.81 &  $14^\circ16{}^\prime12{}^{\prime\prime}$ &  $-0^\circ19{}^\prime12{}^{\prime\prime}$ &        199.15 &                 303.08 &                 130.85 &               Y \\
8  &               39.82 &  $13^\circ55{}^\prime12{}^{\prime\prime}$ &  $-0^\circ18{}^\prime00{}^{\prime\prime}$ &        239.44 &                 299.28 &                 191.57 &               Y \\
9  &               40.63 &  $13^\circ52{}^\prime48{}^{\prime\prime}$ &  $-0^\circ36{}^\prime36{}^{\prime\prime}$ &        115.05 &                 162.97 &                  81.22 &                 \\
10 &               41.70 &  $14^\circ33{}^\prime36{}^{\prime\prime}$ &  $-0^\circ25{}^\prime12{}^{\prime\prime}$ &        158.51 &                 226.20 &                 111.07 &               Y \\
11 &               45.36 &  $13^\circ40{}^\prime12{}^{\prime\prime}$ &  $-0^\circ37{}^\prime48{}^{\prime\prime}$ &        106.74 &                 166.27 &                  68.53 &               Y \\
\bottomrule
\end{tabular}}
\vskip10pt
\begin{tabnote}
$^{a}$ ID of the identified structures.

$^{b}$ The mean velocity of the structures.

$^{c}$ The intensity peak position of the structure in the longitude direction.

$^{d}$ The intensity peak position of the structure in the latitude direction.

$^{e}$ The size  of the structure (Geometric mean of major$\_$radius${}^{f}$ and minor$\_$radius${}^{g}$).

$^{f}$ Major radius of the projection onto the position-position (PP) plane, computed from the intensity weighted second moment in direction of greatest elongation in the PP plane.

$^{g}$ Minor radius of the projection onto the position-position (PP) plane, computed from the intensity weighted second moment perpendicular to the major axis in the PP plane.

$^{h}$ "Y" indicates clouds extending outside the mapped area.
\end{tabnote}
\label{tab:large-size structure}
\end{table*}

\begin{table}
\tbl{Properties of clouds. }{
\begin{tabular}{cccccccc}
\toprule
No $^{a}$ & v$_{\rm cen}$ $^{b}$ & R$\_$major $^{c}$ & R$\_$minor $^{d}$ &           mass $^{e}$ & v$_{\rm rms}$ $^{f}$ & boundary $^{g}$ \\
{} &          (km s$^{-1}$) &                  (pc) &                  (pc) & (10$^{3}$M$_{\odot}$) & \multicolumn{2}{l}{(km s$^{-1}$)} \\
\midrule
1 &                18.79 &                  2.27 &                  0.79 &                  7.48 &                              0.89 &                 \\
3 &                20.51 &                  5.85 &                  1.89 &                 37.21 &                           1.16 &      Y \\
5 &                35.03 &                  2.51 &                  1.09 &                  1.33 &                              0.91 &                 \\
6 &                36.21 &                  2.57 &                  1.29 &                  1.65 &             2               1.17 &    Y             \\
\bottomrule
\end{tabular}}
\vskip10pt
\begin{tabnote}
The distances to all the clouds are assumed to be 2kpc.\\
$^{a}$ Number of the identified structures (corresponding to table \ref{tab:large-size structure}).

$^{b}$ The mean velocity of the structures.

$^{c}$ Major radius of the projection onto the position-position (PP) plane, computed from the intensity weighted second moment in direction of greatest elongation in the PP plane.

$^{d}$ Minor radius of the projection onto the position-position (PP) plane, computed from the intensity weighted second moment perpendicular to the major axis in the PP plane.

$^{e}$ Mass of the cloud.

$^{f}$ Intensity-weighted second moment of velocity.

$^{g}$ "Y" indicates clouds extending outside the mapped area.
\end{tabnote}
\label{tab:focusing_structure}
\end{table}

\newpage

\begin{figure*}[hbtp]
\begin{center}
\includegraphics[width=80mm]{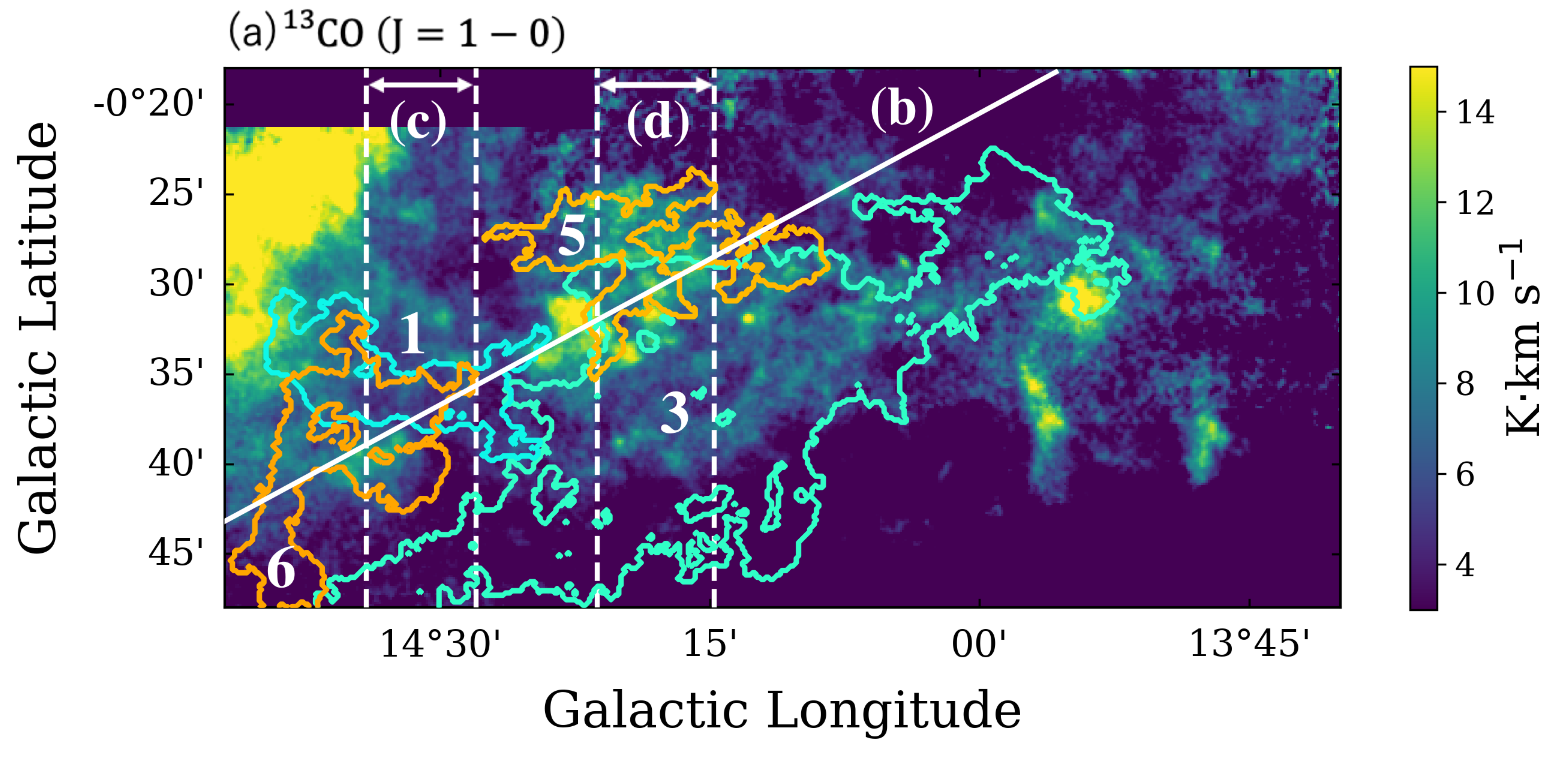}
\includegraphics[width=80mm]{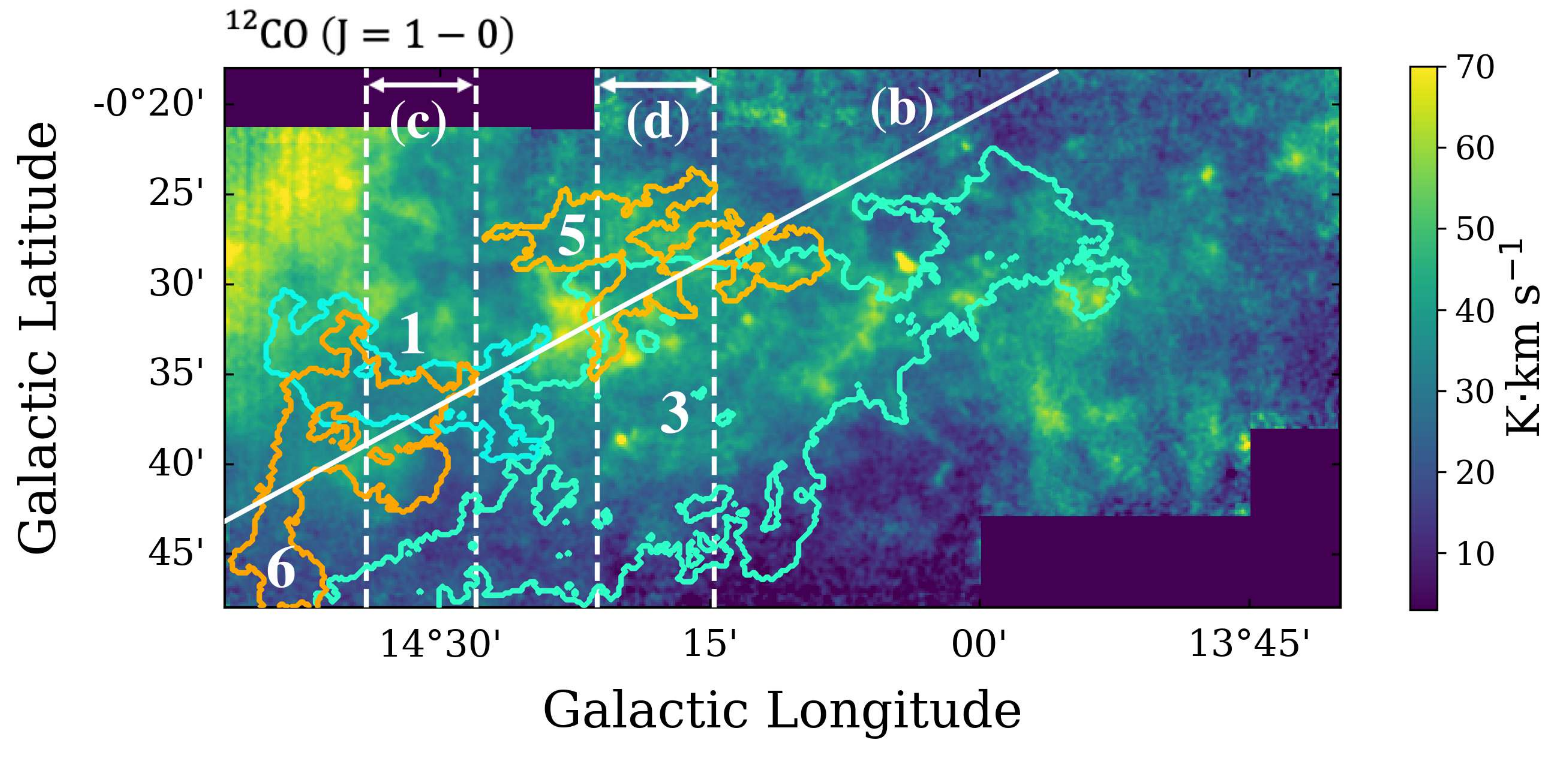}
\includegraphics[width=80mm]{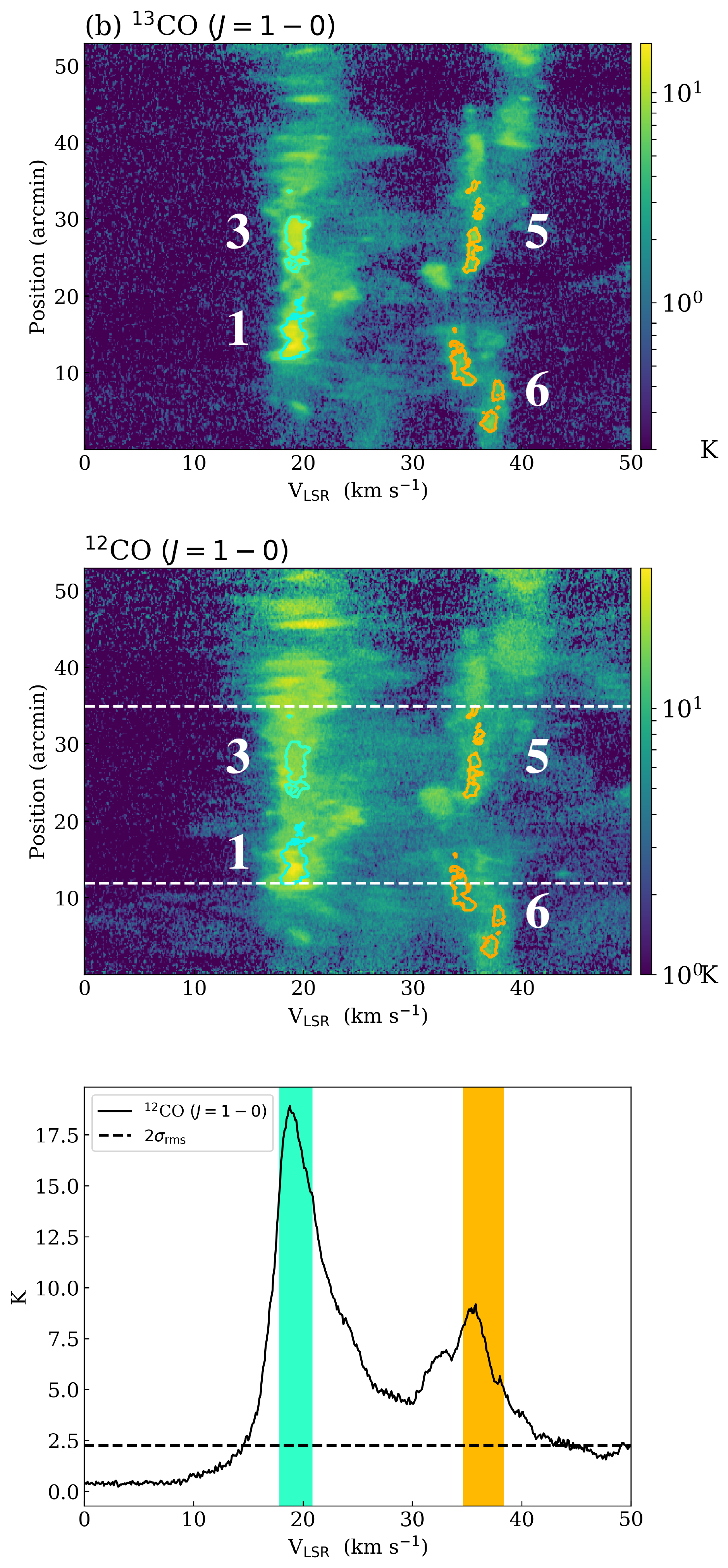}
\end{center}
\end{figure*}

\begin{figure*}[hbtp]
\begin{center}
\includegraphics[width=80mm]{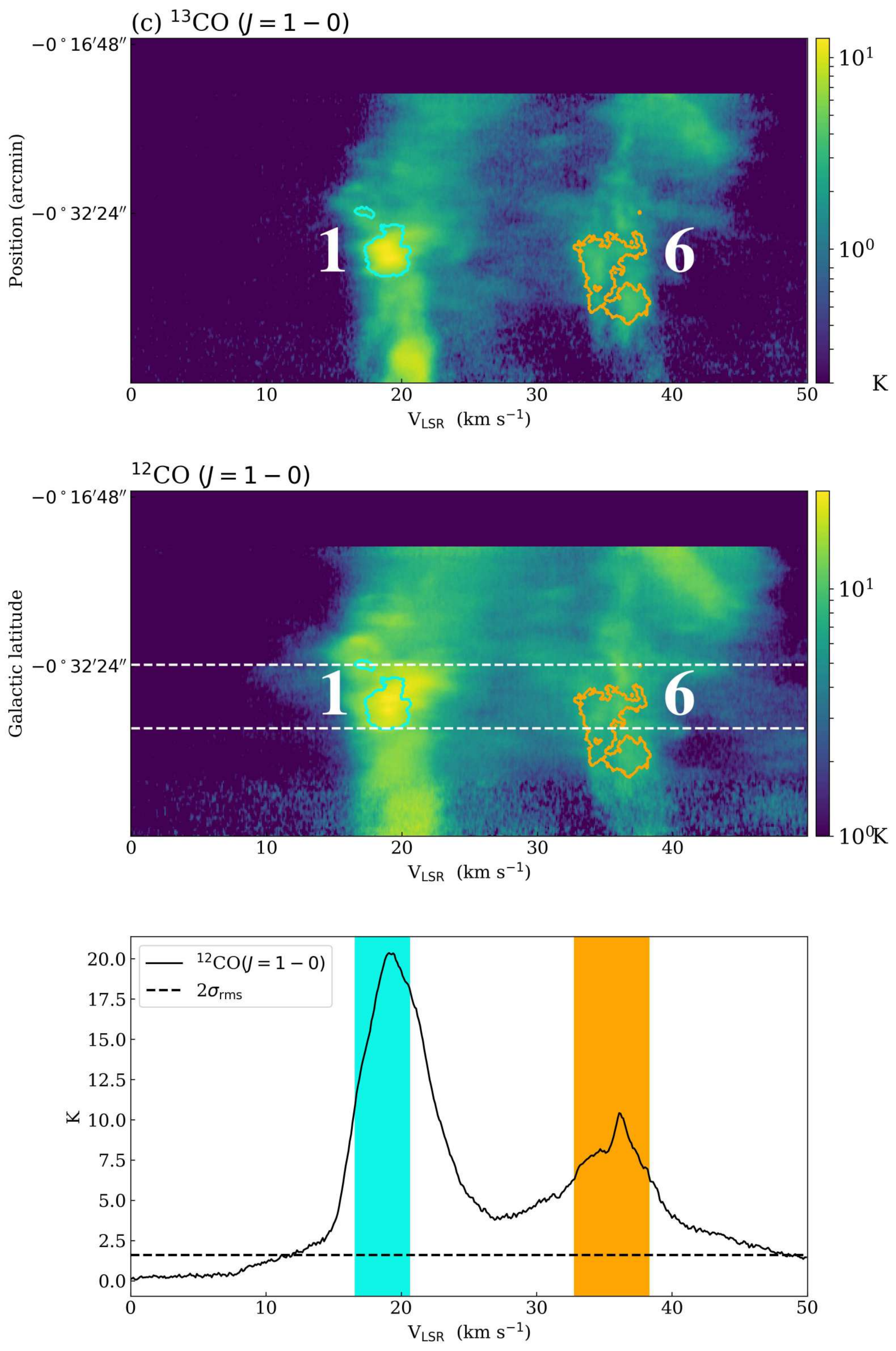}
\includegraphics[width=80mm]{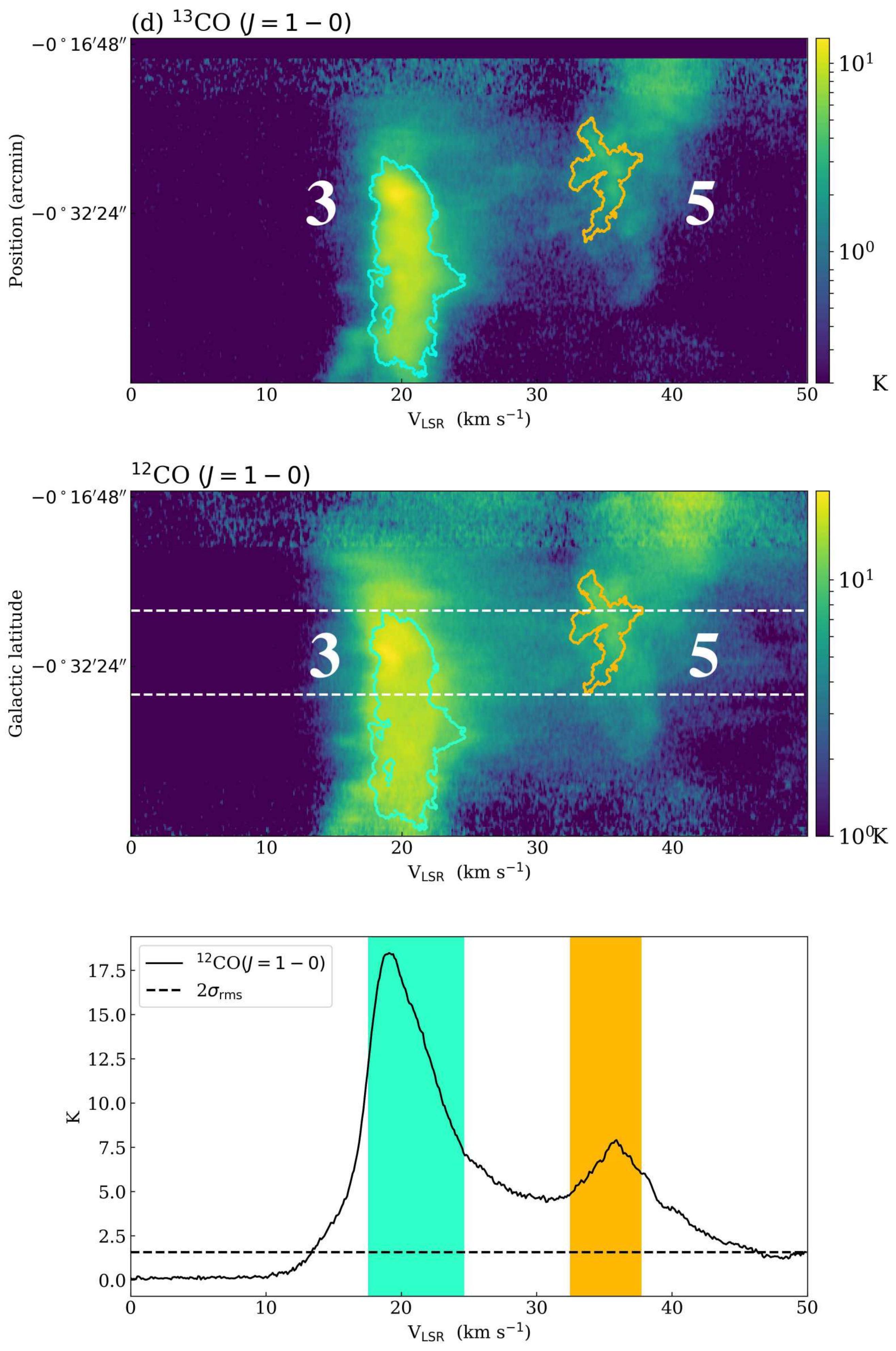}
\begin{flushleft}
\caption{
(a) Integrated intensity maps of the $^{13}$CO$(J=1-0)$ and $^{12}$CO$(J=1-0)$ emission lines. Colored contours represent some $^{13}$CO structures identified by SCIMES (see figure \ref{fig:scimes}) . The labels denote the ID in table \ref{tab:focusing_structure}. The velocity range used for the integration is 24.6 km s$^{-1}$ $<V_{\rm LSR}<32.5$ km s$^{-1}$, which is between the velocity ranges of the No1/No3 and No5/No6 structures.

(b) The upper and middle panels show P-V diagrams of the $^{13}$CO$(J=1-0)$ and $^{12}$CO$(J=1-0)$ emission lines taken along the white solid line in panel (a). The bottom panel shows the mean $^{12}$CO spectra of the region within the two white dashed lines in the middle panel. Velocity ranges of the identified clouds are shaded in their respective colors. The 2$\sigma_{\rm rms}$ level is indicated, where $\sigma_{\rm rms}$ is the average rms noise level of the $^{12}$CO$(J=1-0)$ data within the range used for mean spectra.

(c and d) same as (b) except for the respective regions as indicated by dashed white lines in panel (a).

}
\label{fig:p-v_12CO}
\end{flushleft}
\end{center}
\end{figure*}

\begin{figure*}[hbtp]
\begin{center}
\includegraphics[width=150mm]{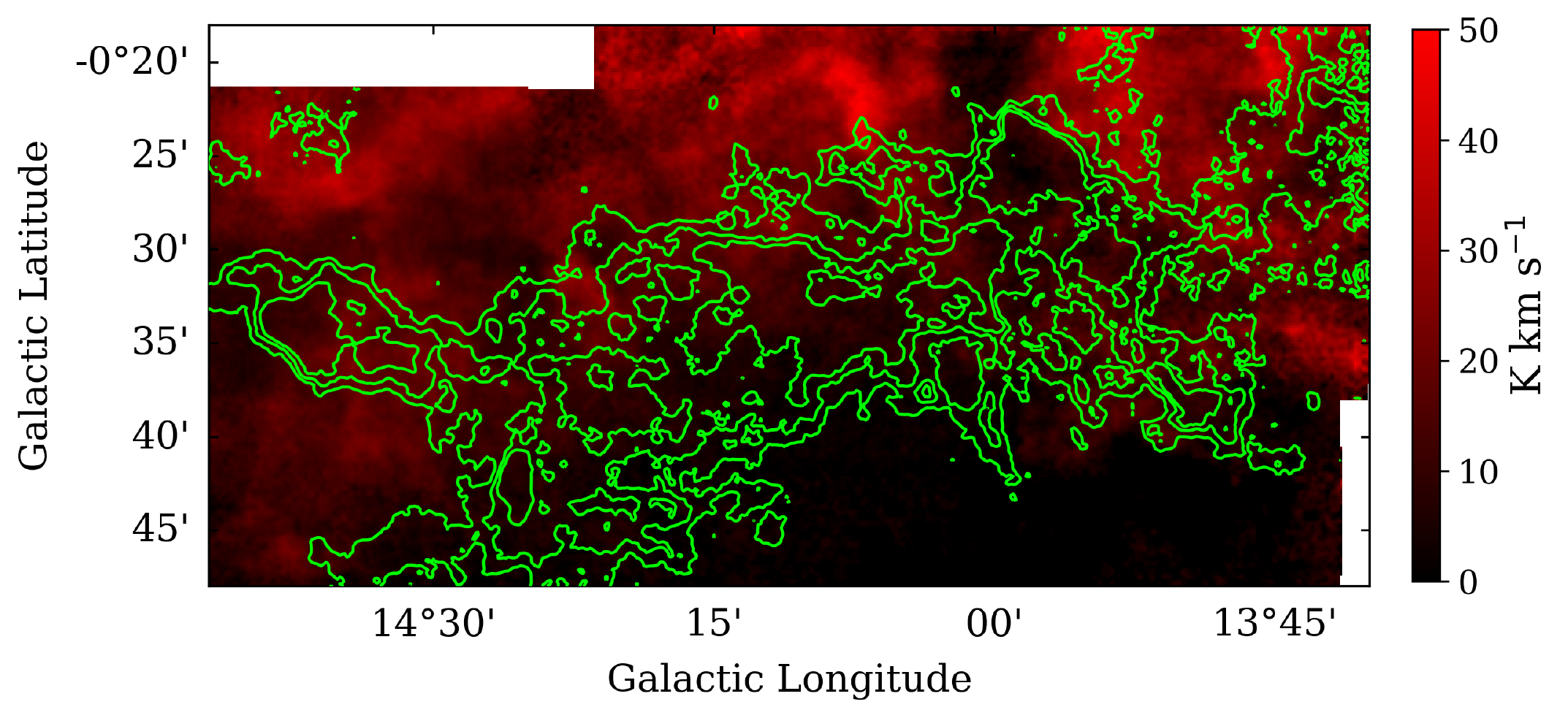}
\begin{flushleft}
\caption{The $^{13}$CO(J=1$-$0) intensity integrated from 10 km s$^{-1}$ to
30 km s$^{-1}$ (green contours) and from 30 km s$^{-1}$ to 50 km s$^{-1}$ (red map). Contours are drawn at 30, 40 and 50 K km s$^{-1}$. Distributions of  the $^{13}$CO  components in the 10$-$30 km $^{-1}$ and 30-50 km $^{-1}$ ranges show an anticorrelation. (see also \citet{Sugitani_2019})
}
\label{fig:mixed_intensity-map}
\end{flushleft}
\end{center}
\end{figure*}


\begin{figure*}[H]
\begin{center}
\includegraphics[width=150mm]{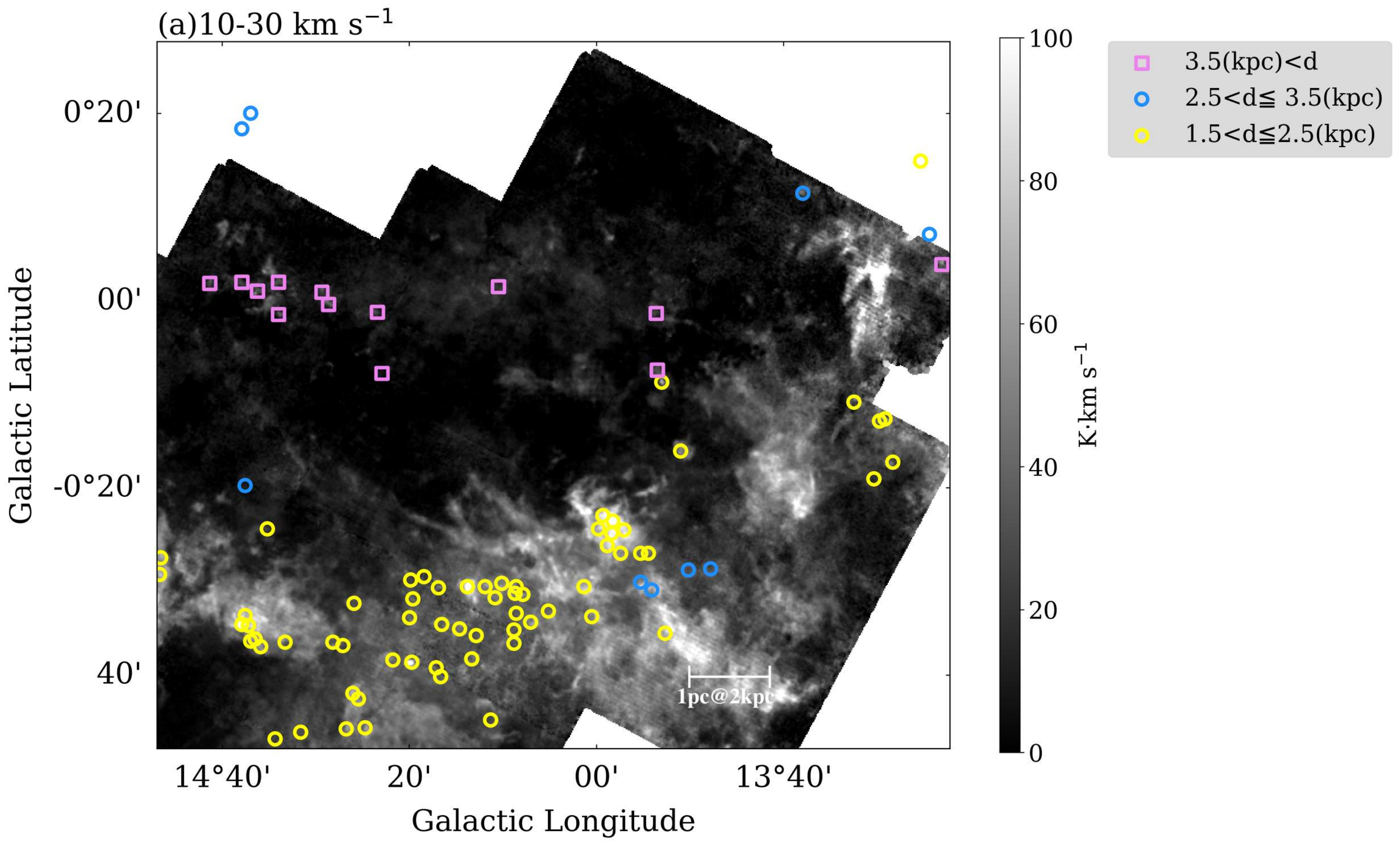}
\vskip15pt
\includegraphics[width=150mm]{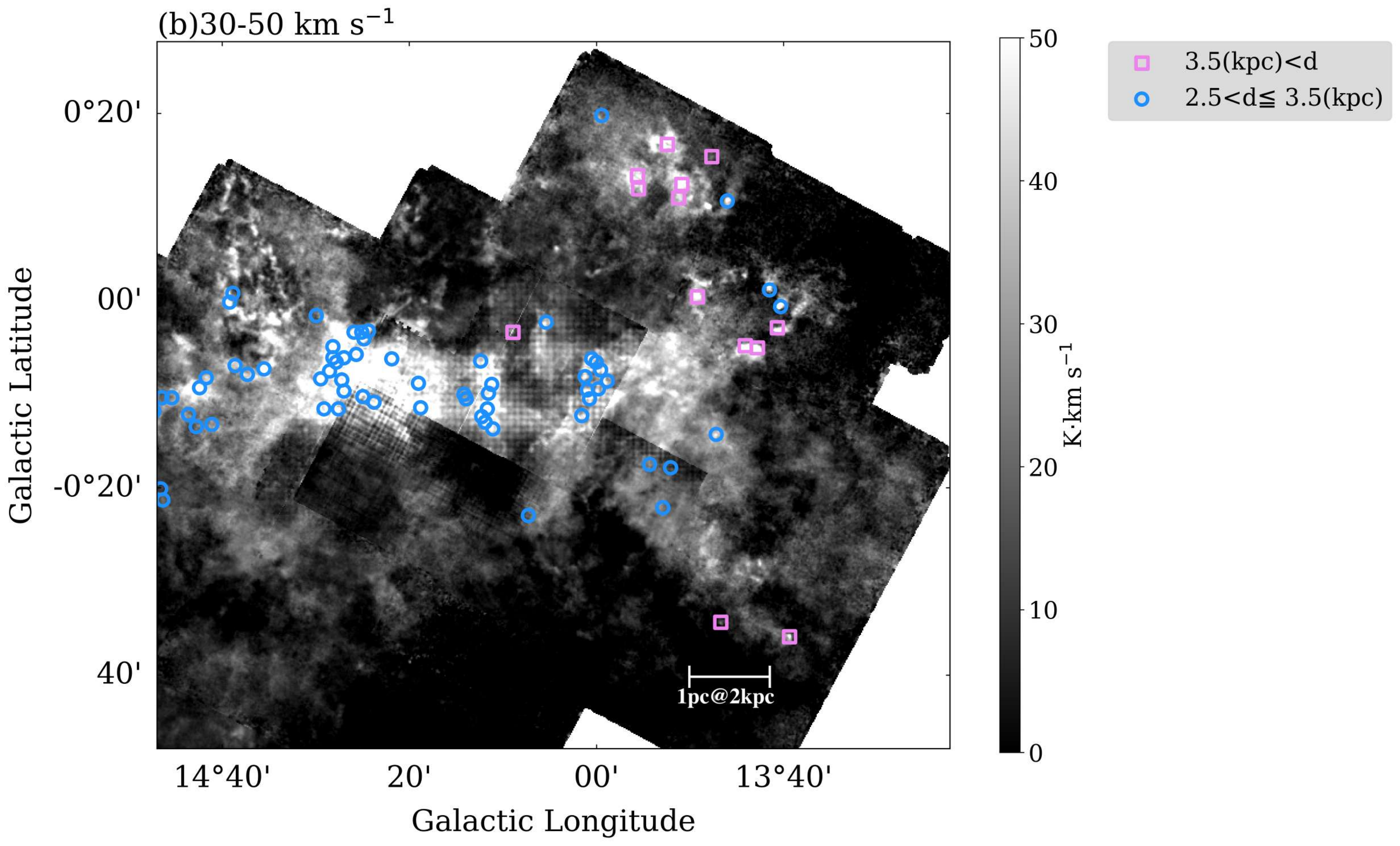}
\end{center}

\begin{center}
\begin{flushleft}
\caption{Integrated intensity maps of the $^{12}$CO$(J=3-2)$ emission line. The velocity ranges used for the integration are (a) 10km s$^{-1}$ $<V_{\rm LSR}<30$ km s$^{-1}$ and (b) 30 km s$^{-1}$ $<V_{\rm LSR}<50$ km s$^{-1}$. In both panels, we plot clumps cataloged by \citet{Urquhart2018}, which fall in the velocity ranges used for the integration, by symbols with different colors. Shapes and colors of each symbol represent the distances.}
\label{fig:12CO(J=3-2)map}
\end{flushleft}
\end{center}
\end{figure*}

\begin{figure*}[hbtp]
\begin{center}
\includegraphics[width=120mm]{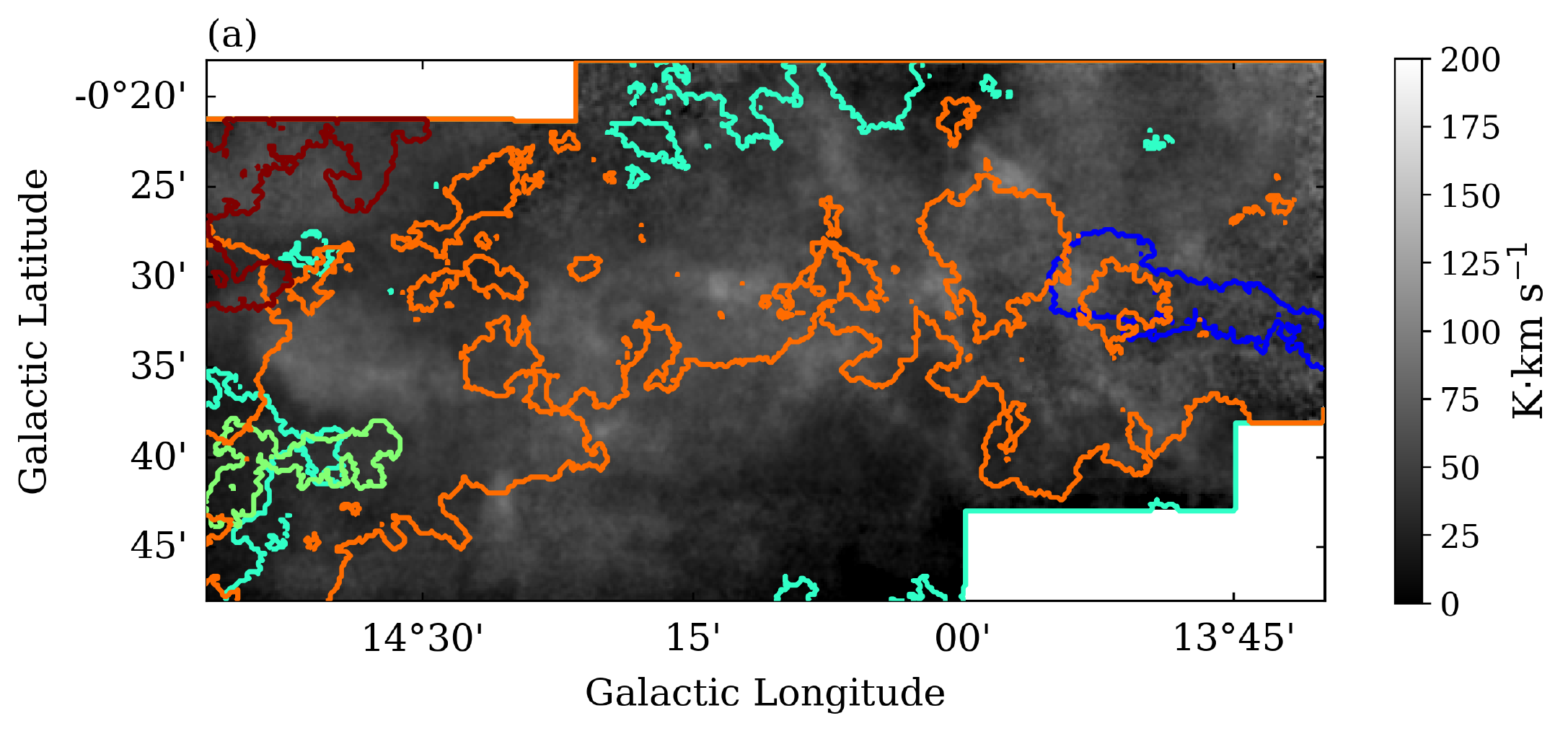}
\includegraphics[width=110mm]{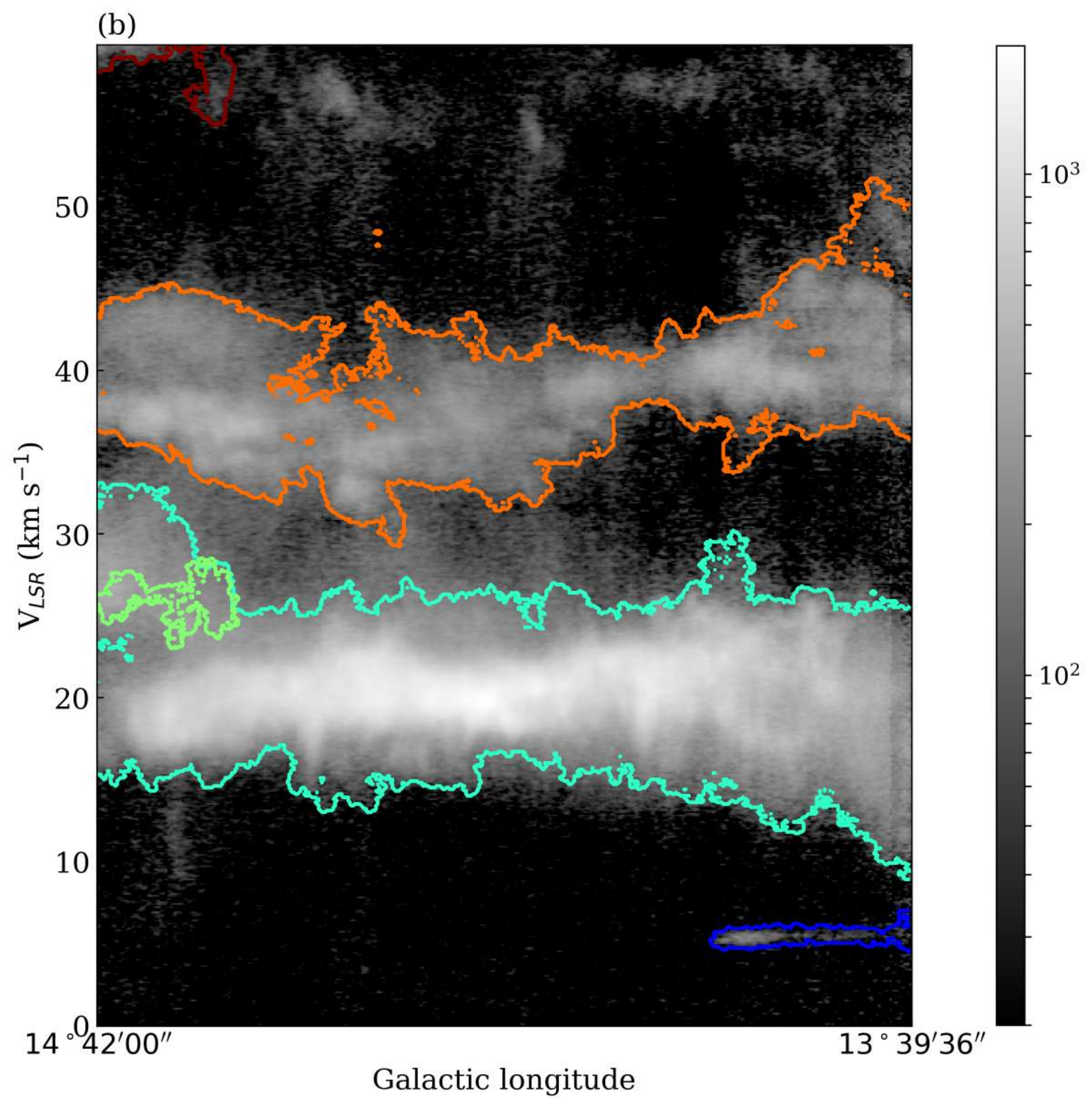}

\begin{flushleft} 
\caption{Clouds identified by SCIMES using three parameters of {\tt  min$\_$value}=5 $\sigma_{\rm rms}$,{\tt min$\_$delta}=3 $\sigma_{\rm rms}$, and 
{\tt min$\_$npix}=30.

(a) Identified structures whose radii are larger than $1.7^{\prime}$ by SCIMES. Contours represent outline of every structure orthogonally projected onto the plane of the sky. Each contour is colored depending on the mean velocity of the structure, according to the colorbar at the bottom of this panel. The background image is the integrated intensity map of the $^{13}$CO(J=1$-$0) emission. The velocity range used for the integration is $-$18 km s$^{-1}$ $<V_{\rm LSR}<60$ km s$^{-1}$.

(b) Structures identified by SCIMES with radii greater than $1.7^{\prime}$ shown on the longitude-velocity diagram generated by integrating the spectra along the latitude direction. Colors of the contours are the same as in panel (a). The background image is the $^{13}$CO$(J = 1-0)$ data integrated in galactic latitude direction}

\label{fig:scimes-5sigma}
\end{flushleft}
\end{center}
\end{figure*}

\begin{figure*}[hbtp]
\begin{center}
\includegraphics[width=170mm]{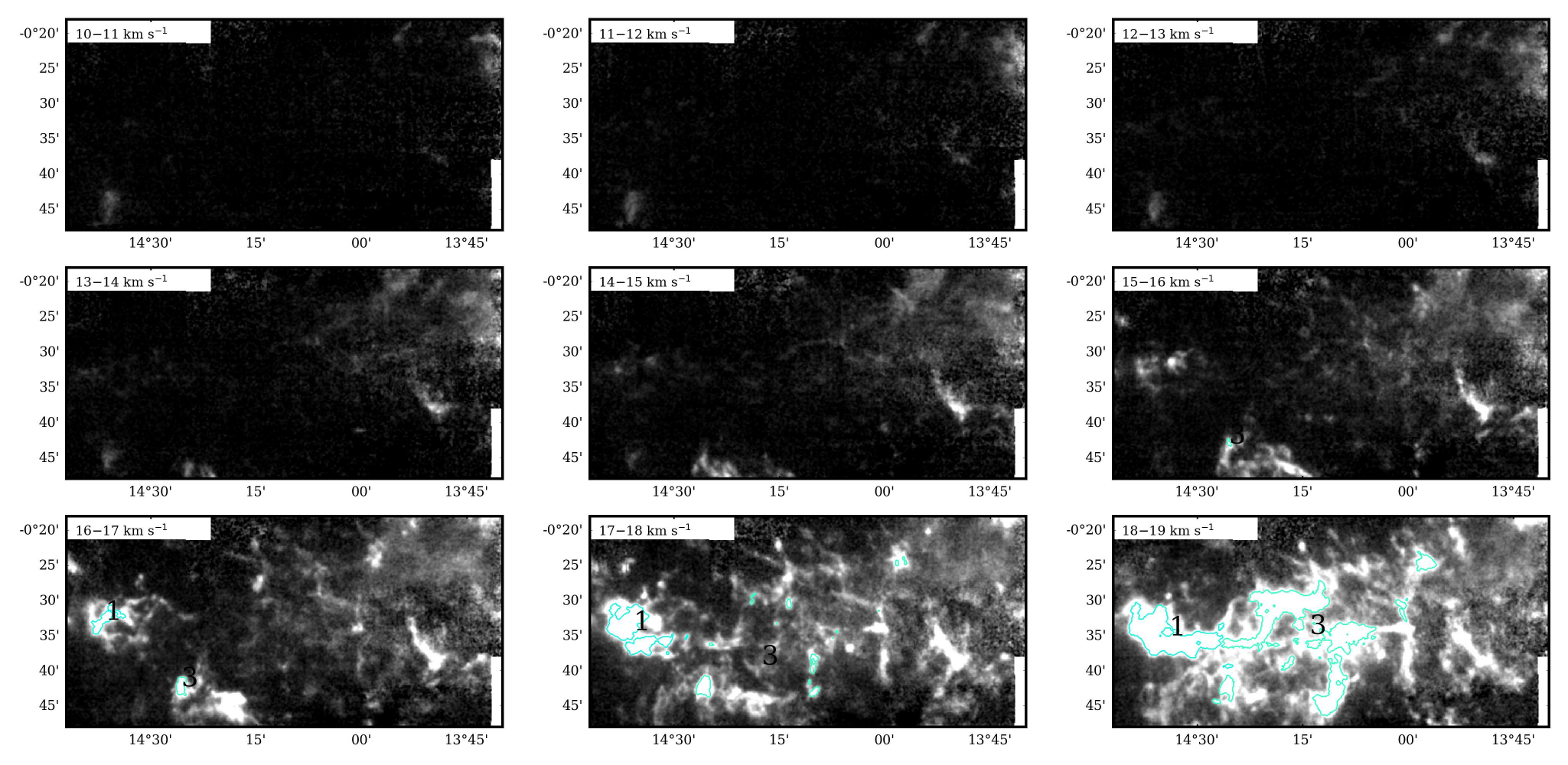}
\includegraphics[width=170mm]{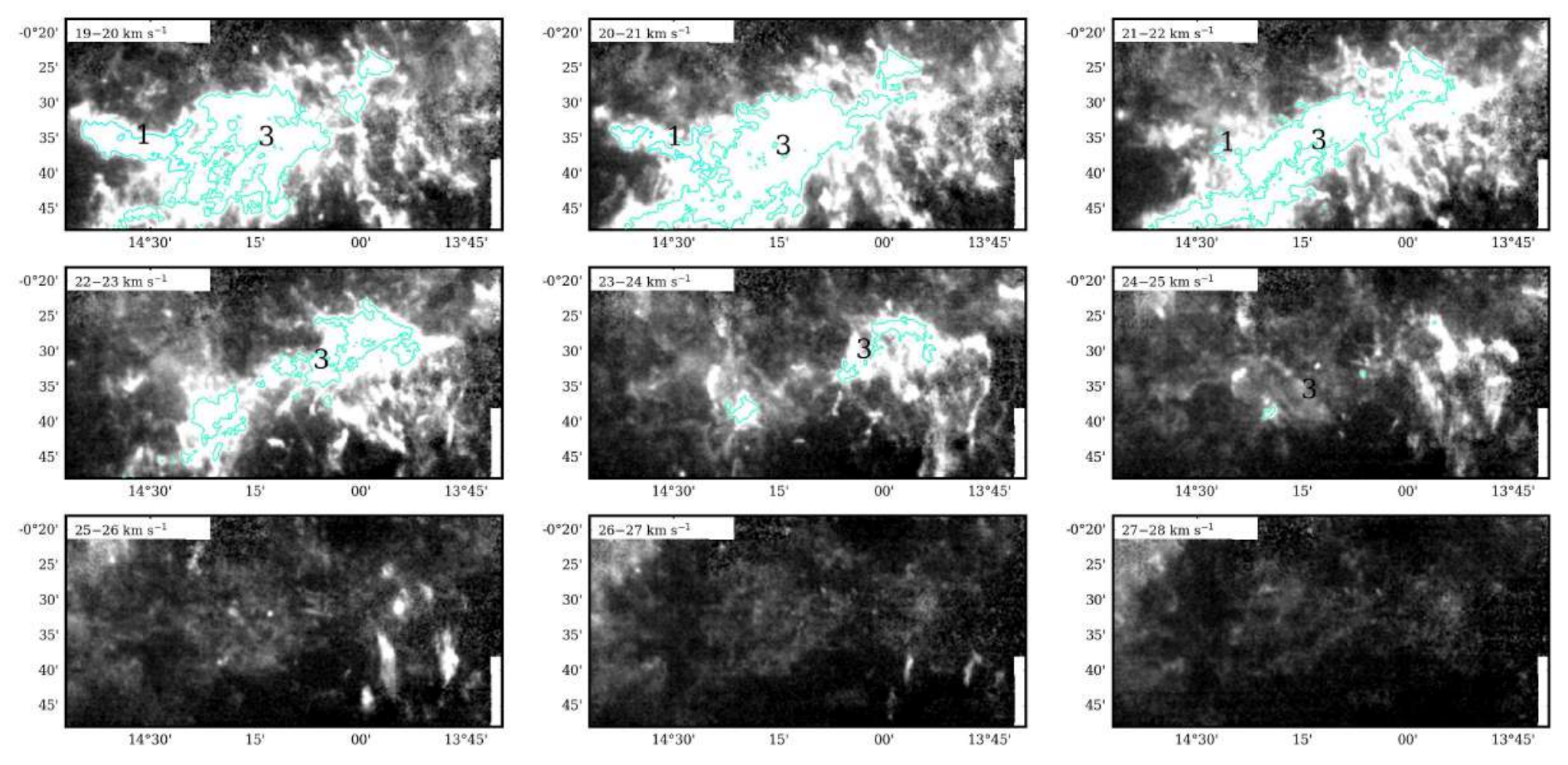}
\includegraphics[width=170mm]{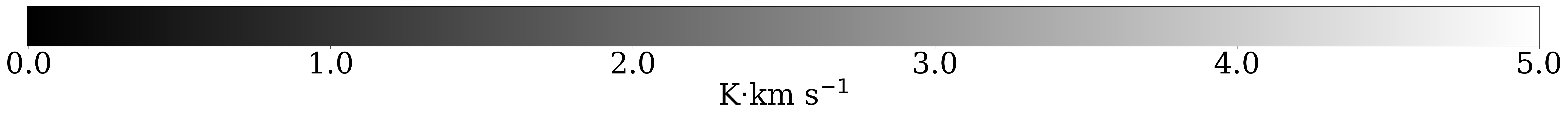}
\end{center}
\end{figure*}

\begin{figure*}[hbtp]
\begin{center}
\includegraphics[width=170mm]{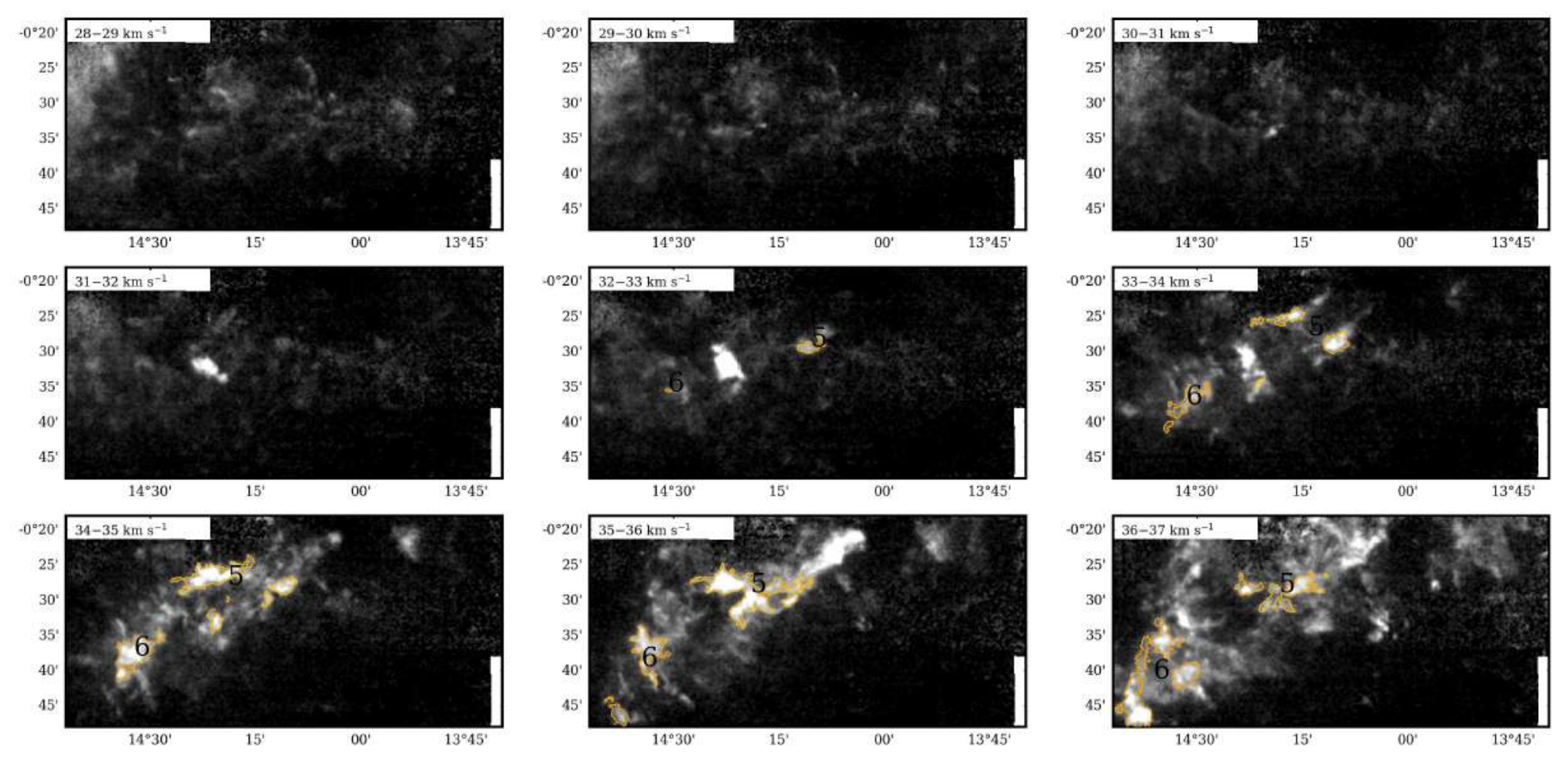}
\includegraphics[width=170mm]{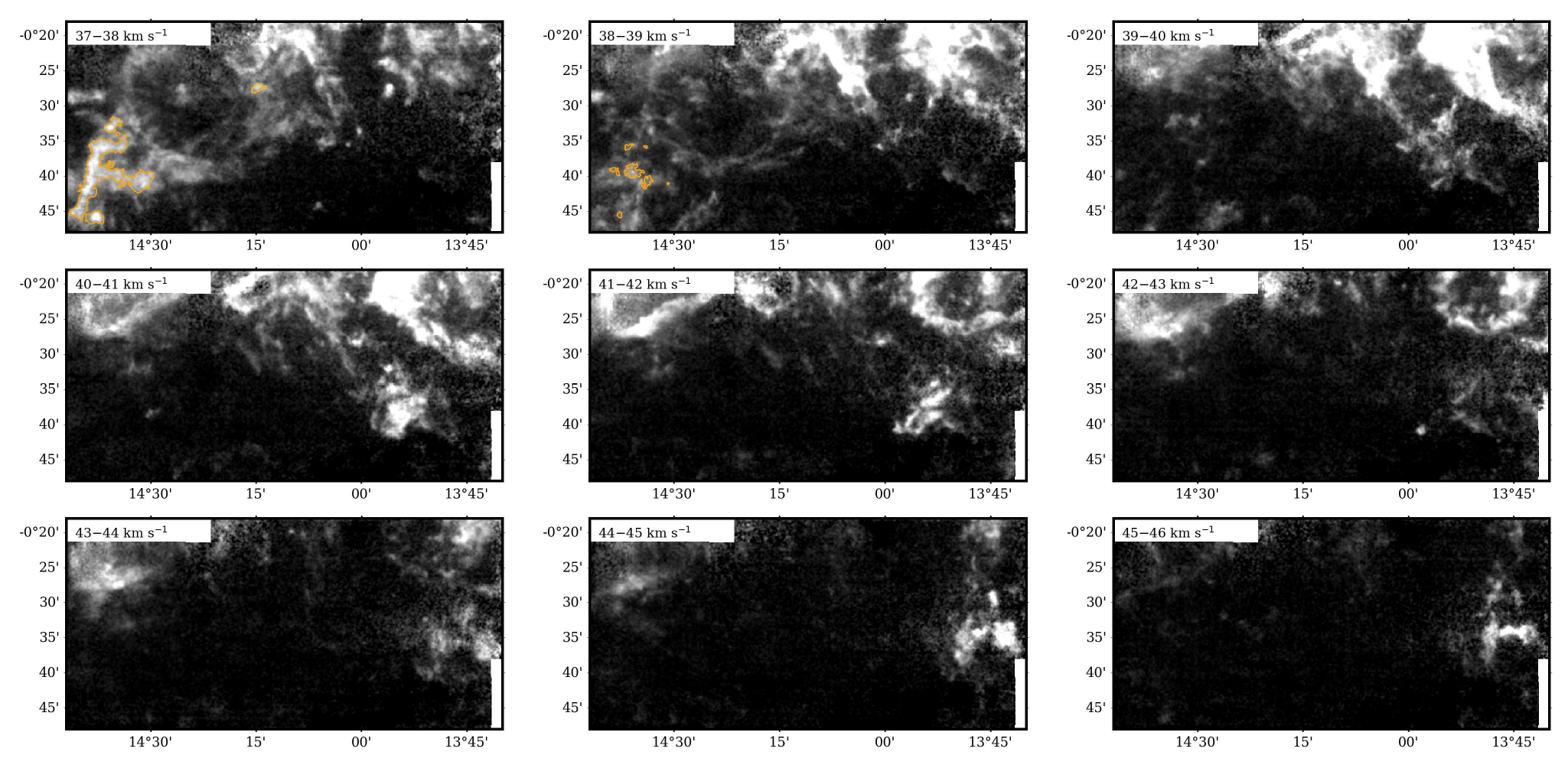}
\includegraphics[width=170mm]{figure_image/velocity-graybar.pdf}
\end{center}
\end{figure*}

\begin{figure*}[T]
\includegraphics[width=170mm]{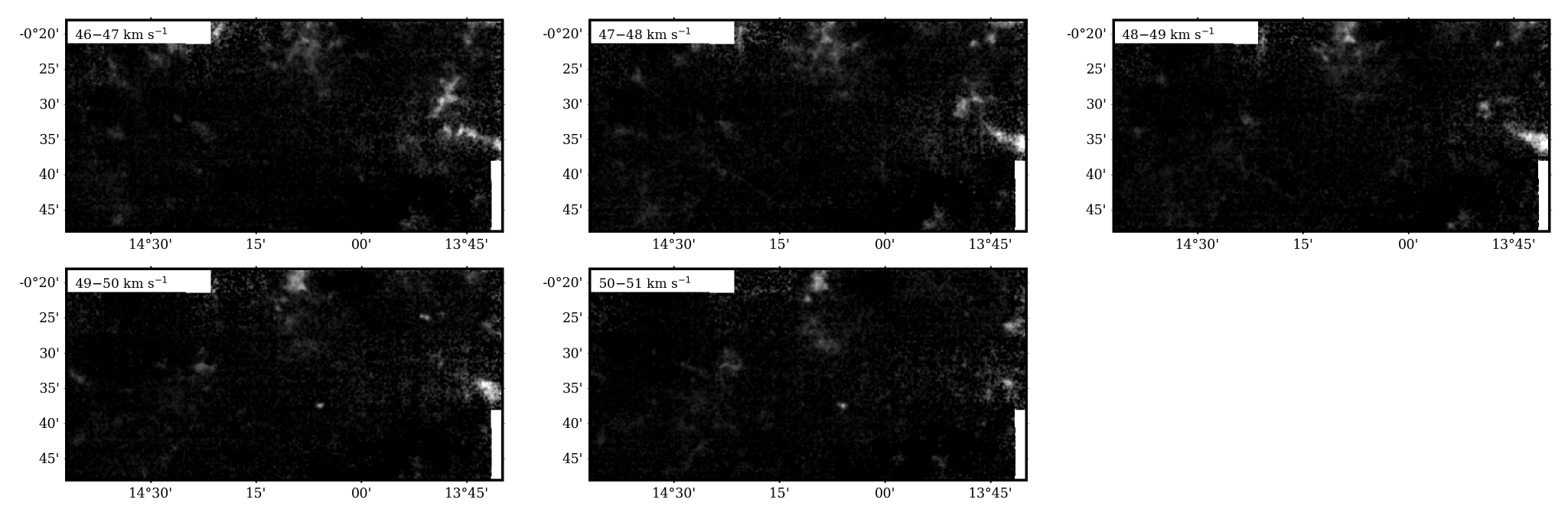}
\includegraphics[width=170mm]{figure_image/velocity-graybar.pdf}
\begin{flushleft}
\caption{Intensity maps of M17 SWex integrated for every 1 km s$^{-1}$ within 10$-$50 km s$^{-1}$. For comparison, the No.1, 3, 5, and 6 clouds are shown by contours in some panels with the corresponding velocities. In the upper-left corner, velocity range used for the integration is shown.}
\label{fig:channelmap}
\end{flushleft}
\end{figure*}

\begin{figure*}[hbtp]
\begin{center}
\includegraphics[width=80mm]{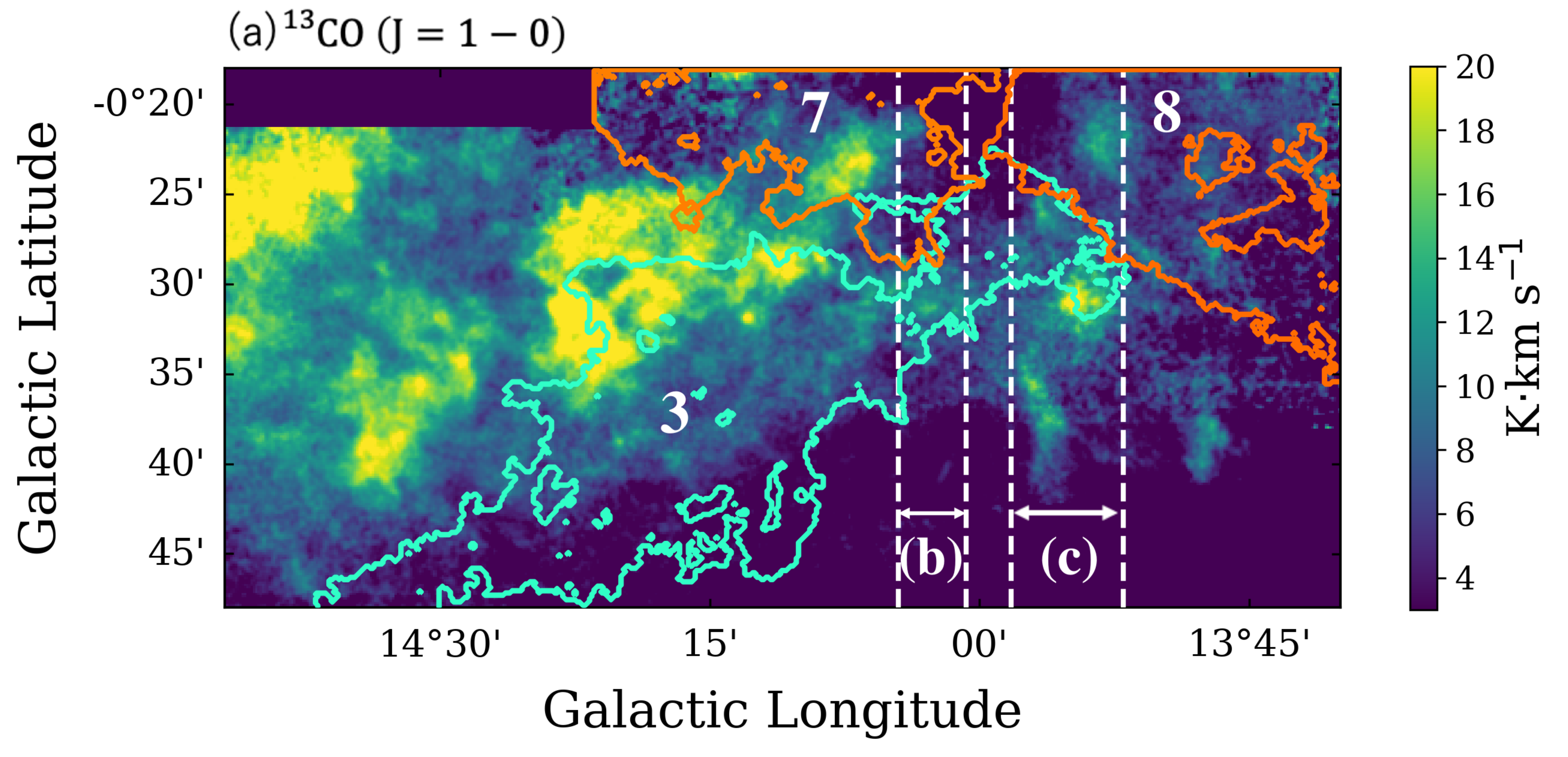}
\includegraphics[width=80mm]{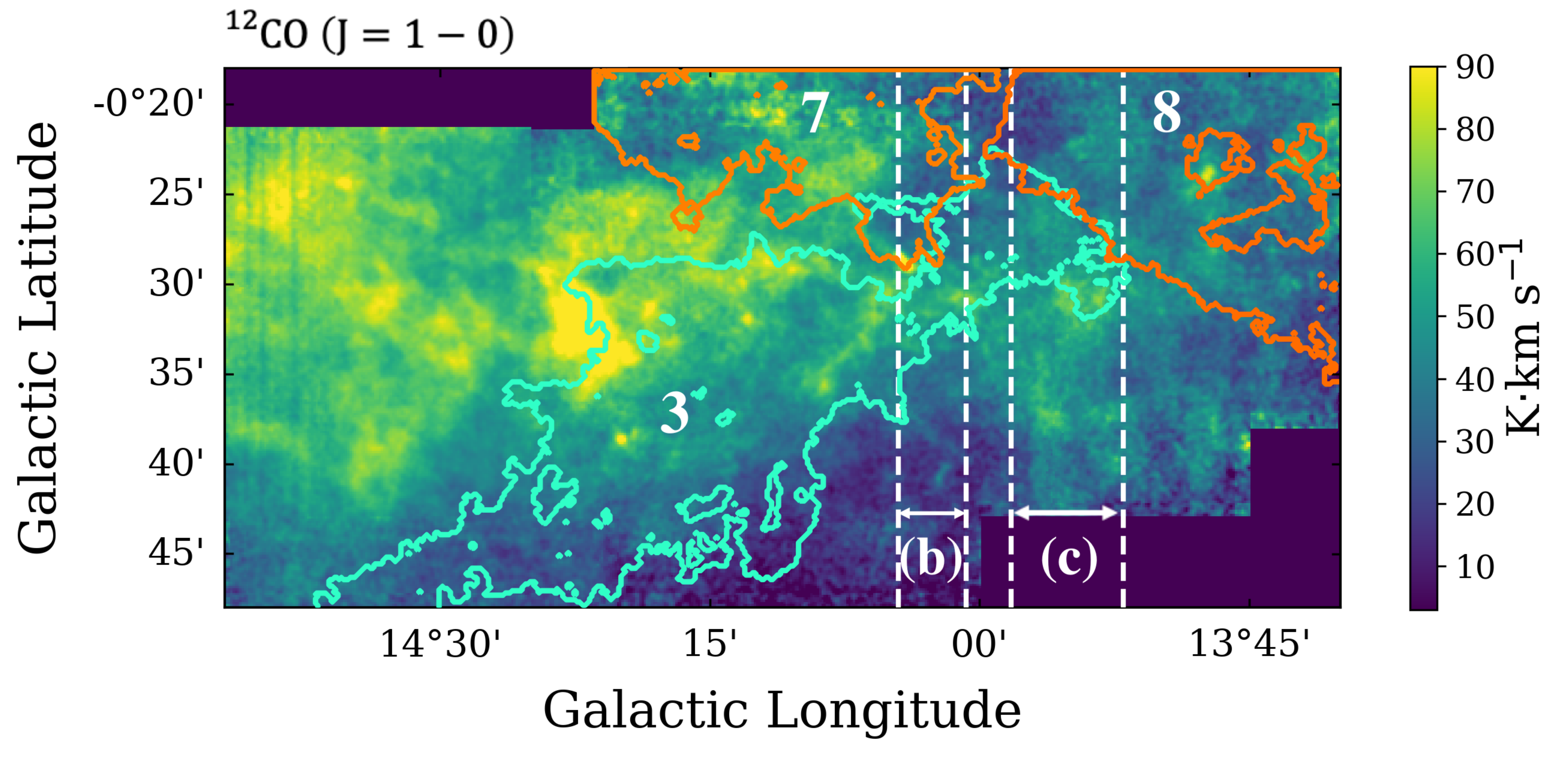}

\end{center}
\end{figure*}

\begin{figure*}[hbtp]
\begin{center}
\includegraphics[width=80mm]{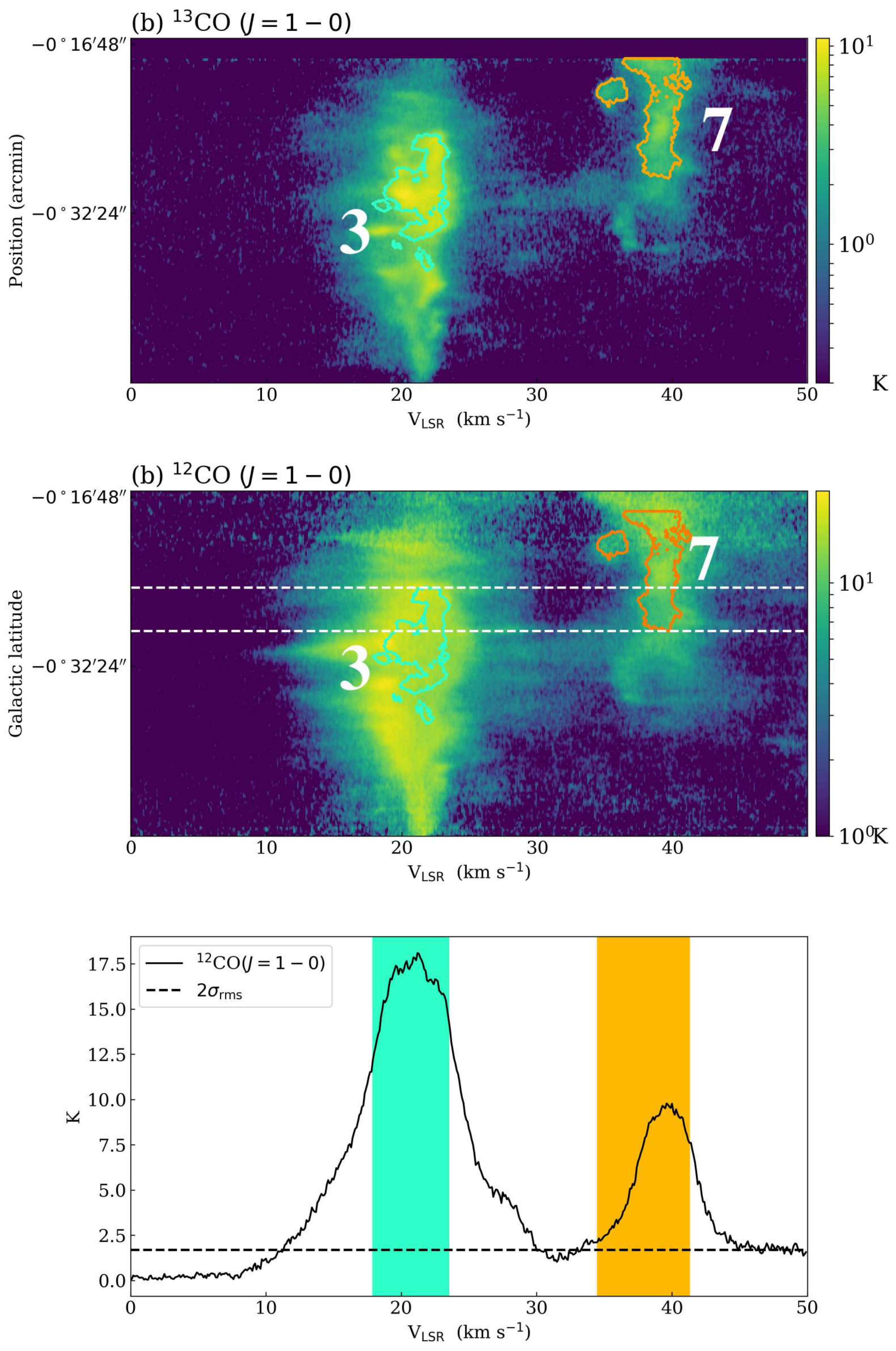}
\includegraphics[width=80mm]{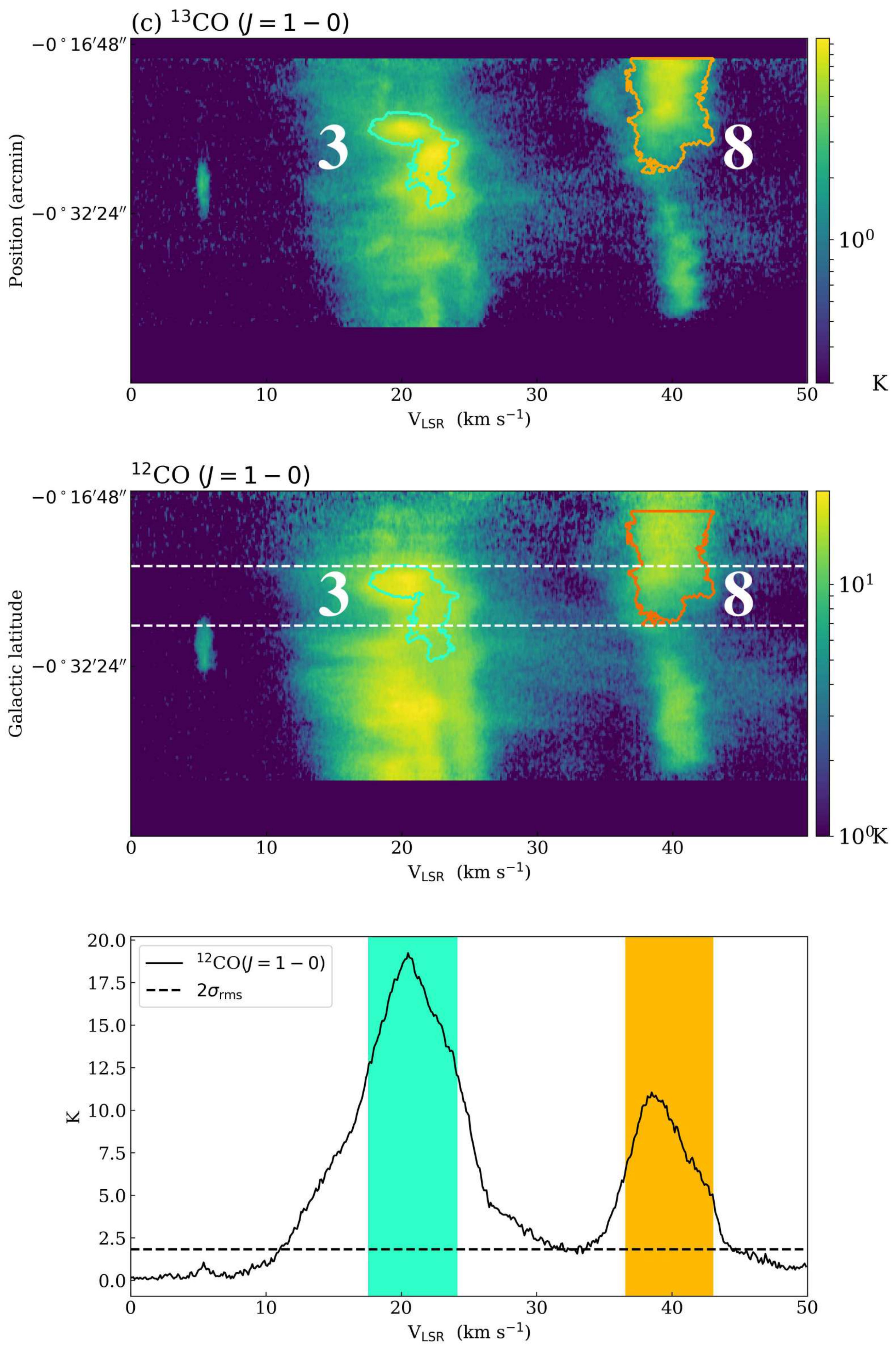}
\begin{flushleft}
\caption{(a) Integrated intensity maps of the $^{13}$CO$(J=1-0)$ and $^{12}$CO$(J=1-0)$ emission lines. Colored contours represent some $^{13}$CO structures identified by SCIMES (see figure \ref{fig:scimes}) . The labels denote the ID in table \ref{tab:focusing_structure}. The velocity range used for the integration is 24.6 km s$^{-1}$ $<V_{\rm LSR}<36.4$ km s$^{-1}$, which is between the velocity ranges of the No1/No3 and No7/No8 structures. 

(b and c) The upper and middle panels shows mean P-V diagram of the $^{13}$CO$(J=1-0)$ and $^{12}$CO$(J=1-0)$ emission lines taken within the range white broken lines indicate in panel (a). The bottom panel shows the mean $^{12}$CO spectra of the region within the two white dashed lines in the middle panel. Velocity ranges of the identified clouds are shaded in their respective colors. The 2$\sigma_{\rm rms}$ level is indicated, where $\sigma_{\rm rms}$ is the average rms noise level of the $^{12}$CO$(J=1-0)$ data within the range used for mean spectra.}

\label{fig:non-bridge}
\end{flushleft}
\end{center}
\end{figure*}

\begin{figure*}[hbtp]
\includegraphics[width=170mm]{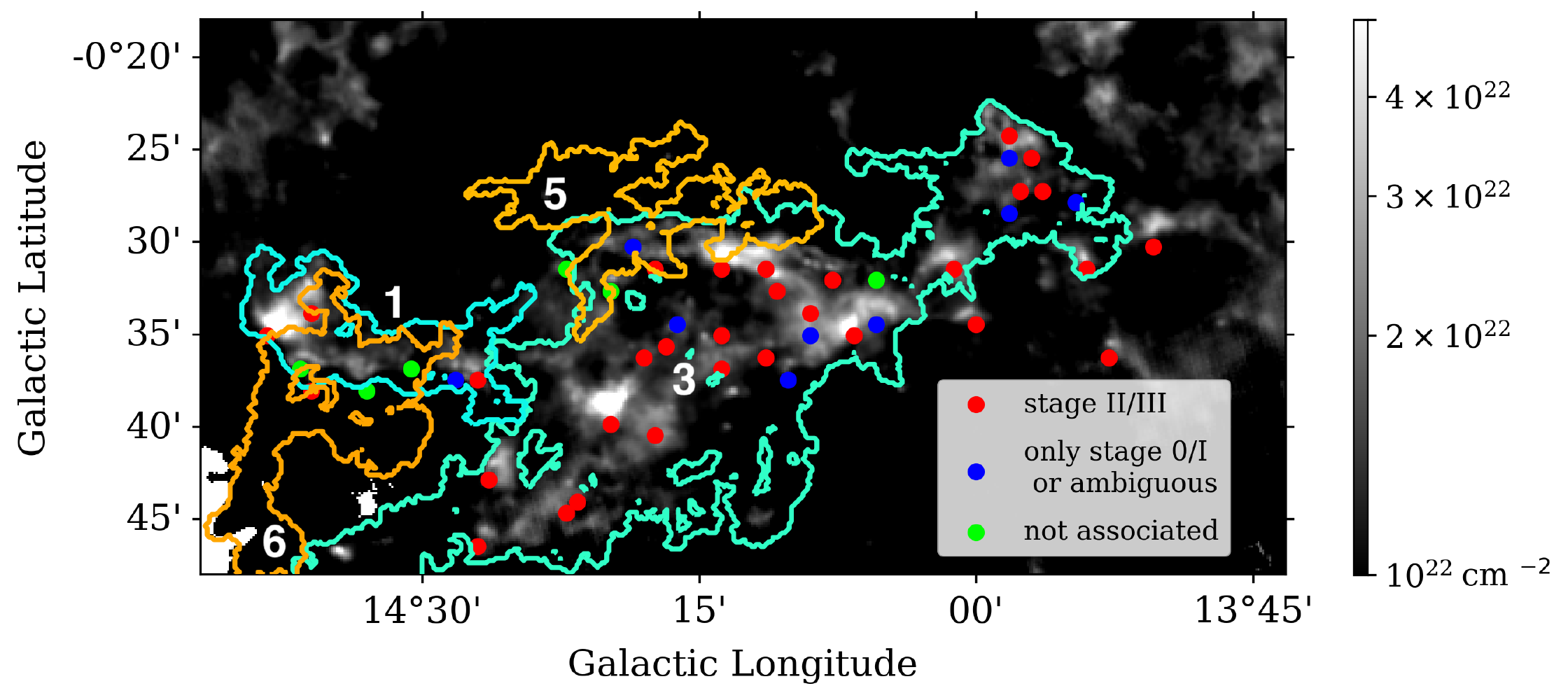}
\begin{flushleft}
\caption{Distribution of the molecular cores identified by \citet{Shimoikura_2019}. Cores associated with stage $\rm{II/III}$ YSOs are shown by the red filled circle. Cores not associated with stage $\rm{II/III}$ but stage $\rm{0/I}$ YSOs or ambiguous stage YSOs are shown by the blue filled circle. Cores not associated with YSOs are shown by the green filled circle. The background is the H$_{2}$ column density map derived from the Herschel data. Colored contours represent the Nos 1,3,5,and 6 cloud structures (see table \ref{tab:focusing_structure}).}
\label{fig:core_column}
\end{flushleft}
\end{figure*}

\end{document}